\documentclass{mn2e}
\usepackage{graphicx}

\def\go{
\mathrel{\raise.3ex\hbox{$>$}\mkern-14mu\lower0.6ex\hbox{$\sim$}}
}
\def\lo{
\mathrel{\raise.3ex\hbox{$<$}\mkern-14mu\lower0.6ex\hbox{$\sim$}}
}
\def\simeq{
\mathrel{\raise.3ex\hbox{$\sim$}\mkern-14mu\lower0.4ex\hbox{$-$}}
}

\def\etal{{et al.\ }}

\begin{document}

\title[Multi-wavelength observations of GRB 080810]{Multi-wavelength observations of the energetic GRB 080810: detailed mapping of the broadband spectral evolution\thanks{This paper is dedicated to the memory of Professor Martin Turner, who sadly passed away during its writing. Martin was an influential figure in X-ray Astronomy and an excellent PhD supervisor. He will be greatly missed.}}

\author[K.L. Page \etal]{K.L. Page$^{1}$, R. Willingale$^{1}$, E. Bissaldi$^{2}$, A. de Ugarte Postigo$^{3}$, S.T. Holland$^{4,5,6}$,\and  S. McBreen$^{7,2}$, P.T. O'Brien$^{1}$, J.P. Osborne$^{1}$, J.X. Prochaska$^{8}$, E. Rol$^{1,9}$,\and E.S. Rykoff$^{10}$, R.L.C. Starling$^{1}$, N.R. Tanvir$^{1}$, A.J. van der Horst$^{11,12}$, K. Wiersema$^{1}$, \and B. Zhang$^{13}$, F.J. Aceituno$^{14}$, C. Akerlof$^{15}$, A.P. Beardmore$^{1}$, M.S. Briggs$^{16}$, \and D.N. Burrows$^{17}$, A.J. Castro-Tirado$^{14}$, V. Connaughton$^{16}$, P.A. Evans$^{1}$,  J.P.U. Fynbo$^{18}$, \and N. Gehrels$^{4}$, C. Guidorzi$^{19,20}$, A.W. Howard$^{21}$, J.A. Kennea$^{17}$, C. Kouveliotou$^{11}$,\and C. Pagani$^{17}$, R. Preece$^{16}$,  D. Perley$^{21}$, I.A. Steele$^{20}$ and F. Yuan$^{15}$
\\
$^{1}$ Department of Physics \&
  Astronomy, University of Leicester, University Road, Leicester, LE1 7RH, UK\\
$^{2}$ Max-Planck-Institut f{\" u}r Extraterrestrische Physik, 85748 Garching, Germany\\
$^{3}$  European Southern Observatory, Casilla 19001, Santiago 19, Chile\\
$^{4}$ NASA Goddard Space Flight Center, Greenbelt, MD 20771, USA\\
$^{5}$ Universities Space Research Association, 10211 Wincopin Circle, Suite 500, Columbia, MD 21044-3432, USA\\
$^{6}$ CRESST, Code 668.8, 8800 Greenbelt Road, Goddard Space Flight Centre, Greenbelt, MD 20771, USA\\
$^{7}$ School of Physics, University College Dublin, Dublin 4, Ireland
\\
$^{8}$ Department of Astronomy and Astrophysics, UCO/Lick Observatory,
University of California, Santa Cruz, CA 95064, USA
\\
$^{9}$ Astronomical Institute Anton Pannekoek, University of Amsterdam, PO number 94249, 1090 GE, Amsterdam, NL\\
$^{10}$ Physics Department, University of California at Santa Barbara, 2233B Broida Hall, Santa Barbara, CA 93106, USA\\
$^{11}$ NASA Marshall Space Flight Center, NSSTC, 320 Sparkman Drive, Huntsville, Alabama 35805, USA\\
$^{12}$ NASA Postdoctoral Program Fellow\\
$^{13}$ Department of Physics and Astronomy, University of Nevada, Las Vegas,
NV 89154-4002, USA\\
$^{14}$ Instituto de Astrof{\' i}sica de Andaluc{\' i}a (CSIC), Camino bajo de Hu{\' e}tor 50, PO Box 03004, E-18080 Granada, Spain\\
$^{15}$ University of Michigan, Randall Laboratory of Physics, 450 Church Street, Ann Arbor, MI 48109-1040\\
$^{16}$ University of Alabama in Huntsville, NSSTC, 320 Sparkman Drive, Huntsville, Alabama, 35805, USA\\
$^{17}$ Department of Astronomy and Astrophysics, Pennsylvania State University, 525 Davey Lab, University Park, PA 16802, USA\\
$^{18}$ Dark Cosmology Centre, Niels Bohr Institute, University of Copenhagen, Juliane Maries Vej 30, 2100 Copenhagen O, Denmark\\
$^{19}$ Physics Dept., University of Ferrara, via Saragat 1, I-44100, Ferrara, Italy\\ 
$^{20}$ Astrophysics Research Institute, Liverpool John Moores University, Twelve Quays House,
Egerton Wharf, Birkenhead CH41 1LD\\
$^{21}$ Department of Astronomy, University of California, Berkeley, CA 94720-3411, USA\\
}

\label{firstpage}

\maketitle

\clearpage

\begin{abstract}

GRB 080810 was one of the first bursts to trigger both {\it Swift} and the {\it Fermi} Gamma-ray Space Telescope. It was subsequently monitored over the X-ray and UV/optical bands by {\it Swift}, in the optical by {\it ROTSE} and a host of other telescopes and was detected in the radio by the VLA. The redshift of $z$~=~3.355~$\pm$~0.005 was determined by Keck/HIRES and confirmed by RTT150 and NOT. The prompt gamma/X-ray emission, detected over 0.3--10$^{3}$~keV, systematically softens over time, with E$_{\rm peak}$ moving from $\sim$~600~keV at the start to $\sim$~40~keV around 100~s after the trigger; alternatively, this spectral evolution could be identified with the blackbody temperature of a quasithermal model shifting from $\sim$~60~keV to $\sim$~3~keV over the same time interval. The first optical detection was made at 38~s, but the smooth, featureless profile of the full optical coverage implies that this originated from the afterglow component, not the pulsed/flaring prompt emission. 

Broadband optical and X-ray coverage of the afterglow at the start of the final X-ray decay ($\sim$~8~ks) reveals a spectral break between the optical and X-ray bands in the range 10$^{15}$--2~$\times$~10$^{16}$~Hz. The decay profiles of the X-ray and optical bands show that this break initially migrates blueward to this frequency and then subsequently drifts redward to below the optical band by $\sim$~3~$\times$~10$^{5}$~s. 
GRB~080810 was very energetic, with an isotropic energy output for the prompt component of 3~$\times$~10$^{53}$~erg and 1.6 ~$\times$~10$^{52}$~erg for the afterglow; there is no evidence for a jet break in the afterglow up to six days following the burst.

\end{abstract}


\date{Received / Accepted}

\section{Introduction}

Gamma-Ray Bursts (GRBs) emit large amounts of energy across the full
range of the electromagnetic spectrum, so obtaining panchromatic data
allows a more thorough investigation of the processes involved. Even with the
rapid slewing capability of {\it Swift} (Gehrels et al. 2004), few bursts have good,
simultaneous multi-band observations of the prompt emission. On rare occasions, 
a trigger on
a precursor has allowed the XRT and UVOT (X-ray Telescope and
UV/Optical Telescope; Burrows et al. 2005a; Roming et al. 2005) to be on
target for (most of) the main event (e.g. GRB~060124 -- Romano et al. 2006;
  GRB~061121 -- Page et al. 2007), whereas, in about 10\% of {\it Swift}-detected GRBs, the
duration of the burst has been such that the narrow-field instruments have been able
to observe the tail-end of the prompt emission (e.g. GRB~060607A -- Ziaeepour
  et al. 2008; GRB~070616 -- Starling et
  al. 2008). Thus, whenever prompt emission is
detected by more than just gamma-ray instruments, the GRB becomes potentially
more interesting and informative, allowing models to be tested more rigorously.

In the case of GRB~080810, bright, highly variable emission detected by the BAT (Burst Alert
Telescope; Barthelmy et al. 2005) continued for more than 100~s, with the XRT
and UVOT on target and collecting data from $\sim$~80~s (Page et al. 2008a); a
bright source was detected in both the X-ray and optical bands. 
The gamma-ray emission was also observed by the {\it Fermi}\footnote{Formerly known as GLAST -- Gamma-ray Large Area Space Telescope} Gamma-ray Burst Monitor
(Meegan et al. 2008), making GRB~080810 one of the first GRBs to trigger both it and the BAT; Konus-Wind also detected this burst (Sakamoto et al. 2008b). 
Many 
telescopes reported the detection of the optical afterglow (see Page et
al. 2008b for a summary) 
while Prochaska et al. (2008)
announced a tentative redshift of z~=~3.35 using the Keck/HIRES
echelle spectrometer, which was then confirmed by RTT150 (Burenin et al. 2008)
and NOT (Nordic Optical Telescope; de Ugarte Postigo et
al. 2008a). The Very Large
Array also detected the radio afterglow at 8.46~GHz (Chandra \& Frail 2008) 3--4 days after the burst.

Section~\ref{obs} presents the observations and preliminary results, covering the $\gamma$-ray ($\S$\ref{sec:gamma}), X-ray ($\S$\ref{sec:x}) and optical ($\S$\ref{sec:opt}) bands. Section~\ref{disc} discusses the redshift determination ($\S$\ref{zdet}), using a thermal interpretation to provide an alternative spectral fit ($\S$\ref{therm}) and the multi-wavelength, broadband modelling ($\S$\ref{multi}). The conclusions are given in Section~\ref{conc}. Throughout this paper we follow the convention of F$_{\nu,
  t}$~$\propto$~$\nu^{-\beta}$~t$^{-\alpha}$ (photon spectral index
  $\Gamma$~=~$\beta$+1), where F$_{\nu,
  t}$ is the flux density, $\nu$ -- the observed frequency and t -- the time since the onset of the burst. Errors are given at 90\% confidence unless otherwise stated.

\section{Observations and Analyses}
\label{obs}

{\it Swift} and {\it Fermi} both triggered on GRB~080810 at 13:10:12 UT on 2008 August 10 (this time is used as T$_0$ throughout the paper), with
the {\it Swift}-XRT and UVOT detecting the afterglow as soon as they were on target. The best {\it Swift}
position is that determined from the UVOT refined analysis (Holland \& Page
2008): R.A.~=~23$^h$47$^m$10.48$^s$, decl.~=~+00$^o$19$'$11.3$''$ (J2000;
estimated uncertainty of 0.6$''$), consistent with the ROTSE-III (Robotic Optical Transient Search Experiment; Rykoff 2008)
and NOT (de Ugarte Postigo et al. 2008a) localisations.

\subsection{Gamma-rays}
\label{sec:gamma}

\begin{figure}
\begin{center}
\includegraphics[clip,width=8.8cm]{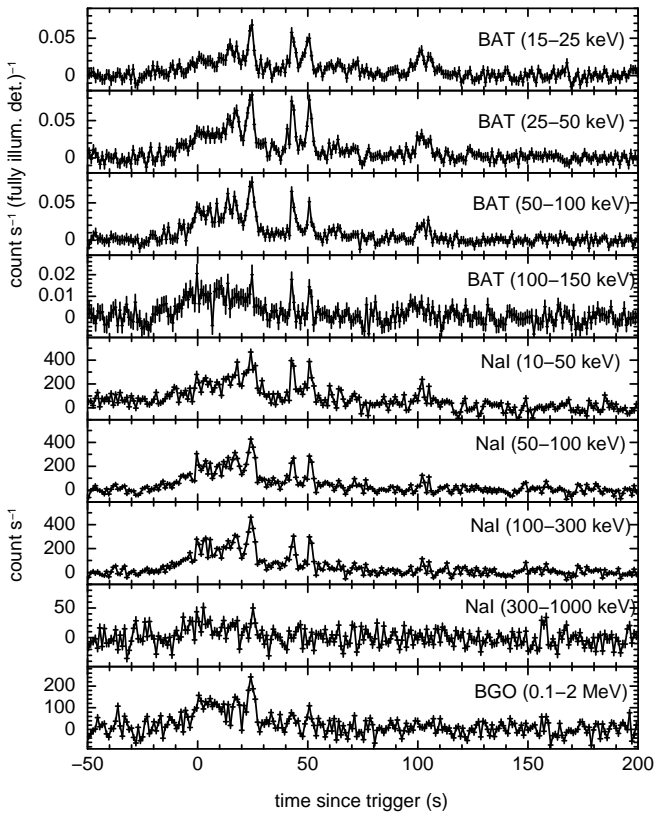}
\caption{The {\it Swift}-BAT (top four panels) and {\it Fermi}-GBM (bottom five panels) light-curves, over their standard energy bands. Note
  that the ordinate scale for the 100--150 keV BAT curve (fourth panel) and the 300-1000~keV NaI curve (eighth panel) are
  different from the lower-energy bands, because the emission was much
  weaker. The BGO light-curve is shown down to 100~keV, but spectral analysis is performed only for data $>$200~keV.
The {\it Swift} light-curves are in units of count s$^{-1}$ (fully illuminated detector)$^{-1}$, while the {\it Fermi} curves are count s$^{-1}$.}
\label{batlc}
\end{center}
\end{figure}

GRB~080810 was clearly detected by the {\it Swift}-BAT (Sakamoto et al. 2008a) over all energy bands
(Figure~\ref{batlc}), although the emission above about 100~keV is
weaker than at the lower energies. The T$_{90}$ (15--150~keV) is 108~$\pm$~5~s (estimated error including
systematics); the fluence over this time is 4.2~$\times$~10$^{-6}$~erg~cm$^{-2}$. The slow rise of the emission, over which there are multiple,
overlapping peaks, started about 20~s (observer's frame) before the trigger.

Konus-Wind also detected GRB~080810, but observed the burst in waiting
mode (Sakamoto et al. 2008b), meaning only 3-channel spectra were
available, covering 20~keV -- 1~MeV. 

The {\it Fermi} Gamma-ray Burst Monitor (GBM; von Kienlin et al. 2004; Meegan et al. 2009) triggered on the burst as well (Meegan et
al. 2008), identifying the same pulses as did the BAT (Figure~\ref{batlc}). Unfortunately the burst
was outside the field of view of the LAT (Large Area Telescope), which is
sensitive to higher energy photons than GBM (0.02--300~GeV, compared to
8~keV--40~MeV). 

\begin{table}
\begin{center}
\begin{tabular}{lcc}
\hline
Instrument  & Band (keV) & T$_{90}$ (s)\\
\hline
{\bf Average}\\
{\it Swift}-BAT & 15--150 & 108~$\pm$~4\\
{\it Fermi}-NaI & 10--1000 & 113~$\pm$~2\\
{\it Fermi}-BGO & 200--20 000 & 73~$\pm$~7\\

\hline
{\bf Energy-sliced}\\
{\it Swift}-BAT & 15--100 & 105~$\pm$~4\\
{\it Swift}-BAT & 100--150 & 55~$\pm$~9\\
{\it Fermi}-NaI & 10--50 & 107~$\pm$~1\\
{\it Fermi}-NaI & 50--100 &  81~$\pm$~1\\
{\it Fermi}-NaI & 100--300 & 73~$\pm$~1\\

\hline
\end{tabular}
\caption{T$_{90}$ measurements over a range of energy bands. The longer duration at lower energies shows that the emission softened over time.}
\label{t90}
\end{center}
\end{table}

The {\it Fermi}-GBM NaI detectors provide similar T$_{90}$ estimates to that measured by the BAT, while the higher energy BGO\footnote{Bismuth Germanate scintillators} durations are shorter; values are given in Table~\ref{t90}. This is a consequence of the hard-to-soft evolution (see, e.g., Table~\ref{jointfit}), combined with the different sensitivities of the instruments.

\begin{table*}
\begin{center}
\begin{tabular}{lcccccc}
\hline
Time  & Detectors & Model & $\Gamma$ & E$_{\rm peak}$ & BB kT & $\chi^2$/dof \\
(s since trigger) & & & & (keV) &  (keV) & \\
\hline
0--10 & BAT & PL & 0.92~$\pm$~0.13 & --- &--- & 66/56\\
0--10 & GBM & PL & 1.48~$\pm$~0.04 & --- & --- &582/488\\
0--10 & GBM & CutPL & 1.06$^{+0.13}_{-0.17}$ & 807$^{+1113}_{-470}$ & --- &509/485\\
0--10 & GBM+BAT & PL & 1.43~$~\pm$~0.03& --- & --- & 694/544 \\
0--10 & GBM+BAT & CutPL & 0.95$^{+0.12}_{-0.13}$ & 602$^{+537}_{-252}$ &--- & 580/543 \\
0--10 & GBM+BAT & PL+Therm & 1.63~$\pm$~0.11 & --- & 62~$\pm$~9 & 597/542\\

\hline

10--20& BAT & PL & 1.24~$\pm$~0.10 & 159$^{+738}_{-92}$ (est.) & --- & 30/56\\
10--20 & GBM & PL & 1.56~$~\pm$~0.04 & --- & --- & 568/486\\
10--20 & GBM & CutPL & 1.07$^{+0.15}_{-0.18}$ & 346$^{+378}_{-160}$ & --- & 505/485\\
10--20 & GBM+BAT & PL & 1.53~$~\pm$~0.03 & --- & --- & 622/544\\
10--20 & GBM+BAT & CutPL & 1.09$^{+0.11}_{-0.12}$ & 353$^{+275}_{-139}$ & --- & 536/543  \\
10--20 & GBM+BAT & PL+Therm & 1.69$^{+0.10}_{-0.09}$ & --- & 46$^{+8}_{-8}$ & 553/542\\
\hline

20--27& BAT & PL & 1.23~$\pm$~0.11 & 162$^{+757}_{-99}$ (est.) & --- & 39/56\\
20--27 & GBM & PL & 1.53~$\pm$~0.04 & --- & --- & 538/486\\
20--27 & GBM & CutPL & 1.11$^{+0.13}_{-0.14}$ & 434$^{+425}_{-190}$ & --- & 476/485\\
20--27 & GBM+BAT & PL & 1.50~$\pm$~0.03 & --- & --- & 596/544  \\
20--27 & GBM+BAT & CutPL & 1.12$^{+0.10}_{-0.11}$ & 452$^{+356}_{-176}$ & --- & 516/543 \\
20--27 & GBM+BAT & PL+Therm & 1.64$^{+0.09}_{-0.08}$ & --- & 53$^{+11}_{-9}$ & 538/542\\
\hline

40--53& BAT & PL & 1.60~$\pm$~0.10 &  82$^{+160}_{-32}$ (est.) & --- & 60/56\\
40--53 & GBM & PL & 1.69~$\pm$~0.06 & --- & --- & 620/486\\
40--53 & GBM & CutPL & 1.27$^{+0.23}_{-0.24}$ & 188$^{+464}_{-114}$ & --- & 602/485\\
40--53 & GBM+BAT & PL & 1.67~$\pm$~0.05 & --- & --- & 682/544 \\
40--53 & GBM+BAT & CutPL & 1.41$^{+0.13}_{-0.16}$ & 230$^{+538}_{-175}$ & -- & 664/543\\
40--53 & GBM+BAT & PL+Therm & 1.78$^{+0.15}_{-0.11}$ & --- & 28$^{+15}_{-9}$ & 668/542\\
\hline

100--106& BAT & PL & 1.71~$\pm$~0.19 & 69$^{+97}_{-52}$ (est.) & --- & 55/56\\
100--106 & GBM & PL & 2.14$^{+0.28}_{-0.23}$ & --- &--- &  600/486\\
100--106 & GBM & CutPL & 0.99$^{+0.83}_{-1.47}$ & 49$^{+472}_{-49}$ & --- & 593/485\\
100--106 & GBM+BAT+XRT & PL & 1.46~$\pm$~0.02 & --- & --- & 904/585\\
100--106 & GBM+BAT+XRT & CutPL & 1.05$^{+0.07}_{-0.08}$ & 39$^{+12}_{-9}$ &--- &  725/584 \\
100--106 & GBM+BAT+XRT & PL+Therm & 1.46$~\pm$~0.03 & --- & 2.6$^{+0.4}_{-0.7}$ & 797/583\\
\hline
\end{tabular}
\caption{Power-law (PL), cut-off power-law (CutPL) and quasithermal (PL+Therm) fits to time-sliced spectra from GBM and BAT; the Fermi NaI n7, n8, nb and BGO b1 detectors were fitted simultaneously each time. The E$_{\rm peak}$ for the BAT single power-law fits are estimated from the relation given by Sakamoto et al. (2009); see text for details.}
\label{jointfit}
\end{center}
\end{table*}

Time-sliced spectra from both the BAT and {\it Fermi} (NaI and BGO) detectors, covering 0--10, 10--20, 20--27, 40--53 and 100--106~s after the trigger, were fitted with single and cut-off power-laws and the results given in Table~\ref{jointfit}; the thermal fits are discussed in Section~\ref{therm}. The useful energy ranges for the BAT, NaI and BGO spectral fitting are 15--150, 8--1000  and 200--40~000~keV, respectively.
The spectra and models were extensively tested in both {\sc xspec} (Arnaud 1996) and the {\it Fermi} software package {\sc rmfit} (Mallozzi et al. 2005); these methods provided consistent results and so the numbers given in this paper are those from {\sc xspec}. Because BAT spectra are created already background-subtracted and have non-Poissonian errors, Cash/Castor statistics cannot be used\footnote{http://heasarc.gsfc.nasa.gov/docs/xanadu/xspec/manual/\\XSappendixCash.html}; hence all results were obtained using $\chi^2$ statistics.

Using the F-test, the Band function (Band et al. 1993) is not a statistical improvement over the simpler cut-off power-law, with $\beta$, the higher energy index, unconstrained in each case. For the fits presented here, the normalisations of the GBM detectors were tied together at a value of 1.23 relative to the normalisation of the BAT, which was itself fixed at unity. This constant of normalisation for the GBM was determined by simultaneously fitting all five intervals of data, but allowing the other fit parameters to vary between the intervals.
The cross-calibration between {\it Swift}-BAT and {\it Fermi}-GBM is discussed in Section~\ref{inter}.

Sakamoto et al. (2009) found a correlation between the photon index from a simple power-law fit to a BAT GRB spectrum and E$_{\rm peak}$, thus allowing an estimate of the peak energy from the limited BAT energy bandpass. The correlation for a source 15$^o$ off-axis (GRB~080810 was approximately 20$^o$ off-axis, so this is the closest of the relationships), log(E$_{\rm peak}$)~=~3.184~$-$~0.793$\Gamma$ (where 1.3~$\leq$~$\Gamma$~$\leq$~2.3), was used to produce the estimated E$_{\rm peak}$ values given in Table~\ref{jointfit} (marked as `est.'); BAT slewed during the interval 12--64~s after the trigger, so E$_{\rm peak}$ was estimated for the last (100--106~s) spectrum using the on-axis approximation.
The spectrum extracted for 0--10~s after the trigger has too hard a photon index to allow the use of this approximation, while 10--20~s is just consistent with the range. The 1$\sigma$ spread of the relation has been included in the error estimation. These estimated peak energies are consistent with those found from jointly fitting the {\it Fermi} and {\it Swift} data, although the error bars on the measurements are very large.

\begin{figure}
\begin{center}
\includegraphics[clip,angle=-90,width=8cm]{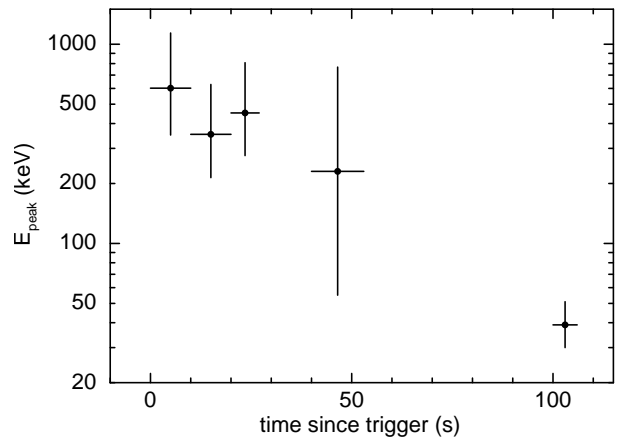}
\caption{E$_{\rm peak}$, measured from joint {\it Fermi}-{\it Swift} fits, moves to lower energies with time.}
\label{epeak}
\end{center}
\end{figure}

The numbers show that the peak energy moves to lower values over time; this is demonstrated graphically in Figure~\ref{epeak}.
The single power-law fits also show that $\Gamma$ increased (softened) until at least 53~s. The spectrum from 100-106~s covers a flare in the XRT emisson, which explains the harder (flatter) photon index (see Figure~\ref{gamma}).

\begin{figure}
\begin{center}
\includegraphics[clip,angle=-90,width=8cm]{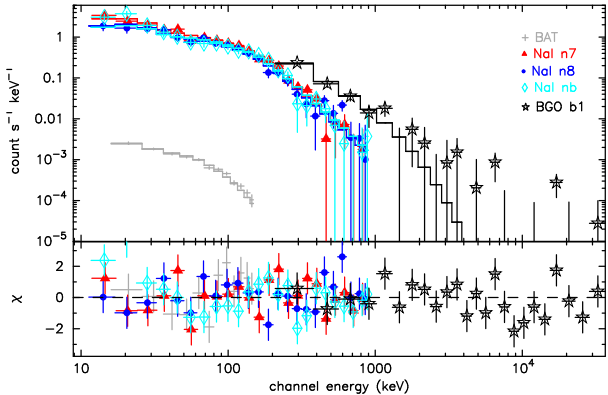}
\caption{GBM and BAT spectra covering -4 to 26~s over the trigger. The residuals are plotted in terms of sigma, with error bars of 1$\sigma$.}
\label{spec}
\end{center}
\end{figure}

Extracting simultaneous BAT and GBM spectra over T$_0$$-$4 -- T$_0$+26~s (i.e., the brightest interval; Figure~\ref{spec}), a cut-off power-law model is significantly better than a single power-law, with $\Gamma$~=~1.05$^{+0.07}_{-0.08}$ and  E$_{\rm peak}$ estimated to be 569$^{+290}_{-181}$~keV, corresponding to an isotropic energy release of E$_{\rm iso}$$\sim$~3~$\times$~10$^{53}$~erg (1~keV -- 10~MeV in the rest frame; z~=~3.355 from Prochaska et al. 2008).

\begin{figure}
\begin{center}
\includegraphics[clip,angle=-90,width=8cm]{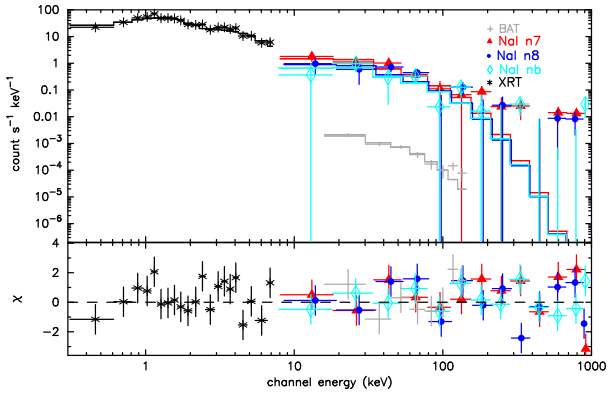}
\caption{{\it Fermi}-GBM, BAT and XRT spectra covering 100--106~s after the trigger. The residuals are plotted in terms of sigma, with error bars of 1$\sigma$.}
\label{allspec}
\end{center}
\end{figure}

Figure~\ref{allspec} plots the {\it Swift}-XRT and BAT and {\it Fermi}-NaI spectra between 100-106~s. There is only a tenuous detection in the BGO at this time, so that spectrum has not been included in the plot for clarity; it was, however, used in the fit to help constrain E$_{peak}$.

Note that, in both of the above plots, the BAT data appear lower down the ordinate axis simply because of the way the normalisations are defined. Inter-calibration between {\it Swift} and {\it Fermi} is discussed in Section~\ref{inter}.


\subsubsection{Inter-calibration}
\label{inter}

The inter-calibration of the {\it Swift}-BAT and the {\it Fermi}-GBM is a work in progress and preliminary simulations were discussed by Stamatikos, Sakamoto \& Band (2008); further results will be presented in Stamatikos et al. (in prep). As mentioned in the previous section, the normalisations of the GBM detectors were all tied together for the current paper, finding a mean value of 1.23~$\pm$~0.06 compared to a BAT value of unity. Allowing the normalisations to vary between the GBM detectors, while again simultaneously fitting all the datasets with a cut-off power-law, produced the following  relative constants: NaI n7~=~1.24$^{+0.09}_{-0.08}$, n8~=~1.18~$\pm$~0.08, nb~=~1.32$^{+0.10}_{-0.09}$ and BGO b1~=~1.88$^{+0.31}_{-0.27}$, where, as before, the BAT constant was fixed to unity. Thus, the GBM detectors agree quite well, with a typical discrepancy of $\lo$~20~per~cent.




\subsection{X-rays}
\label{sec:x}

\begin{figure}
\begin{center}
\includegraphics[clip,angle=-90,width=8cm]{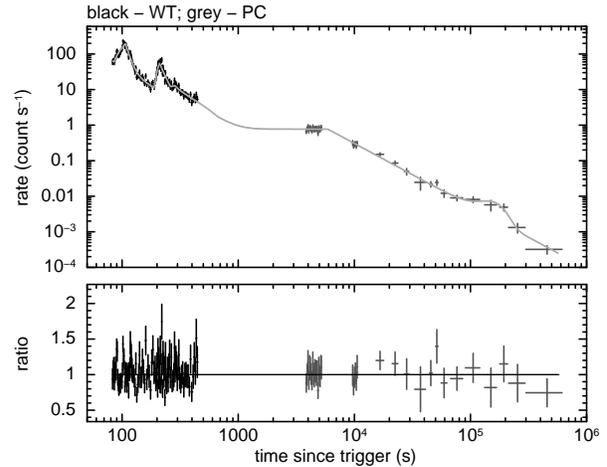}
\caption{0.3--10~keV X-ray light-curve. The decay has been fitted with a
  doubly-broken power-law and overlying flare/Gaussian components (described
  in the text). The lower panel shows the ratio between the data and the model.}
\label{xrtlc}
\end{center}
\end{figure}

{\it Swift}-XRT identified and centroided on an uncatalogued X-ray source in a
2.5~s Image Mode frame, 76~s after the BAT trigger. The source was bright enough such that the XRT stayed
in Windowed Timing (WT) mode throughout the first orbit, which ended about
460~s after the trigger. The WT data showed two large flares, with smaller
peaks superimposed (Figure~\ref{xrtlc}), with an underlying decay of $\alpha$~=1.05$^{+0.17}_{-0.14}$. 

By the time of the second orbit, the source had faded
sufficiently that data could be collected in Photon Counting (PC) mode. The
`canonical' X-ray light-curve identified from {\it Swift} bursts (Nousek et
al. 2006; Zhang et al. 2006) follows a steep-shallow-normal pattern (possibly
with a later jet-break) and, while not all bursts show all these components (Evans et al. 2009),
in this case the first PC orbit could be identified as the end of the shallow
plateau phase. {\it Swift} is in a low-Earth orbit, so can only observe a given target for a maximum of about 1.5--2~ks per
96~minute revolution, leading to orbital gaps in the data. Because of these intervals of no data, the start time of the plateau in the GRB~080810 X-ray light-curve is very
uncertain. Nevertheless, a flattening in the decay at this time was a statistical improvement. The best-fit parameters obtained by fitting the complete light-curve with a
doubly broken power-law (plus flare components, as shown in Figure~\ref{xrtlc})
are given in Table~\ref{power}; because of the limited data, the plateau phase was poorly constrained. There
was a deviation from the late-time power-law decay around $\sim$~150~ks, when the X-ray emission
briefly rebrightened.

The early flares (almost certainly due to the continuation of the prompt emission component) are well fitted with FRED-like (Fast Rise Exponential Decay)
profiles, while the deviation around 10$^{5}$ seconds can be equally well-modelled by either a FRED-like profile or a Gaussian. However, it should be noted that the data are much better sampled at
early times, so the shape of the later flare is more uncertain. 
A model comprising a power-law
decay with six super-imposed, sometimes over-lapping flares (at $\sim$~T$_0$+91, 102,
107, 118, 210 and 281~s; the strongest peaks are those at 102 and 210~s) provides a good fit to the WT data, with each flare
component being statistically significant. There are also additional
deviations, but adding further flares to the model does not improve the
fit.

Including these six burst components at early
times, and a Gaussian to model the later-time rebrightening, we obtain 
$\chi^2$/dof~=~134/130 ($\chi^2_\nu$~=~1.03).

\begin{table}
\begin{center}
\begin{tabular}{lccccc}
\hline
Parameter & Value\\
\hline
$\alpha_{\rm 1}$ & 1.05$^{+0.17}_{-0.14}$\\
T$_{\rm b,1}$ & 2679$^{+1677}_{-1335}$ s\\
$\alpha_{\rm 2}$ & $-$0.03$^{+0.57}_{-0.58}$\\
T$_{\rm b,2}$ & 5909$^{+969}_{-518}$ s\\
$\alpha_{\rm 3}$ & 1.76$^{+0.10}_{-0.08}$\\
   
\hline
\end{tabular}
\caption{The doubly broken power-law model fitted to the decay continuum; the `plateau' phase is poorly constrained. A
  number of flare components are also required to give a good fit. The three
  main early flares peak at 107, 210 and 281~s; a later rebrightening occurs
  around 150~ks. See text for more details.}
\label{power}
\end{center}
\end{table}


The spectrum of the plateau orbit ($\sim$~T$_0$+4--5~ks) can be
modelled with a power-law of $\Gamma$~=~1.95~$\pm$~0.07 absorbed by the Galactic
column of N$_{\rm H}$~=~3.28~$\times$~10$^{20}$~cm$^{-2}$. After the temporal
break, the photon index is $\Gamma$~=~2.00~$\pm$~0.09, so there is no
sign of spectral evolution at this point. A spectrum extracted just for the
late-time `bump' in the light-curve is also consistent with these values, with
$\Gamma$~=~1.99$^{+0.28}_{-0.24}$. The lack of measured intrinsic absorption
is not unusual for bursts at medium to high redshift (see, e.g., Grupe et
al. 2007) and is also consistent with the optical data (see Section~\ref{zdet}).

\subsection{Optical}
\label{sec:opt}

\begin{table}
\begin{center}
\begin{tabular}{lcccc|}
\hline
Filter & Mag. & Time & Exp. time & Telescope\\
 & & (s since burst) & (s)\\
\hline

I & 19.36 $\pm$ 0.09 & 47863 &  300 & 1.5m-OSN\\
I & 19.34 $\pm$ 0.08 & 50548 &  300 & 1.5m-OSN\\
I & 19.42 $\pm$ 0.13 & 52581 &  300 & 1.5m-OSN\\
I & 19.51 $\pm$ 0.08 & 53558 &  300 & 1.5m-OSN\\

i   &        19.91 $\pm$ 0.03 &     52186  &    1800& LT\\
i   &        20.67 $\pm$ 0.06 &    124762  &    1800& LT \\
i   &        21.11 $\pm$ 0.06 &    174122  &    1500& FTN\\
i   &        21.08 $\pm$ 0.05 &    177534  &    1800& FTN\\
i   &        21.54 $\pm$ 0.06 &    224122  &    1800& LT \\
i   &        21.60 $\pm$ 0.07 &    242905  &    1800& FTN\\
i   &        21.66 $\pm$ 0.07 &    261896  &    1800& FTN\\
i   &        21.90 $\pm$ 0.09 &    311386  &     900& INT/WFC \\
i   &        22.63 $\pm$ 0.10 &    401674  &    1200& INT/WFC \\

R   &        15.95 $\pm$ 0.12  &     3320   &    300 & FTS\\
R   &        16.30 $\pm$ 0.12 &      3697  &     180& FTS \\
R   &        16.27 $\pm$ 0.12 &      3886  &     180& FTS \\
R   &        16.46 $\pm$ 0.12 &      4075  &     180& FTS \\
R   &        16.48 $\pm$ 0.12 &      4263  &     180& FTS \\
R   &        16.66 $\pm$ 0.12 &      4453  &     180& FTS \\
R   &        16.39 $\pm$ 0.12 &      4712  &     300& FTS \\
R   &        16.97 $\pm$ 0.12 &      5689  &     300& FTS \\
R   &        16.84 $\pm$ 0.12 &      6624  &     300& FTS \\
R   &        16.97 $\pm$ 0.12 &      7422  &     300& FTS \\
R   &        21.55 $\pm$ 0.09 &    244918  &    1800& FTN\\
R   &        21.56 $\pm$ 0.06 &    265404  &    1800& FTN \\
R   &        19.31 $\pm$ 0.07 &     38117  &     120& NOT \\
R   &        19.66 $\pm$ 0.07 &     43110  &     300& 1.5m-OSN\\
R   &        19.78 $\pm$ 0.04 &     47548  &     300& 1.5m-OSN \\
R   &        19.81 $\pm$ 0.06 &     48212  &     300&  1.5m-OSN\\
R   &        19.75 $\pm$ 0.04 &     48868  &     300& 1.5m-OSN \\
R   &        19.78 $\pm$ 0.06 &     50016  &     300& 1.5m-OSN \\
R   &        19.81 $\pm$ 0.06 &     50215  &     300& 1.5m-OSN \\
R   &        19.76 $\pm$ 0.05 &     50884  &     300&  1.5m-OSN\\
R   &        19.83 $\pm$ 0.05 &     51559  &     300&  1.5m-OSN\\
R   &        19.85 $\pm$ 0.06 &     52255  &     300&  1.5m-OSN\\
R   &        19.89 $\pm$ 0.06 &     52916  &     300&  1.5m-OSN \\
R   &        19.83 $\pm$ 0.08 &     53879  &     300&  1.5m-OSN \\
R   &        19.65 $\pm$ 0.06 &     54530  &     300& 1.5m-OSN  \\
R   &        20.76 $\pm$ 0.09 &    140260  &     900& 1.5m-OSN \\
R   &        22.44 $\pm$ 0.31 &    419360  &    3000& 1.5m-DK \\
R   &        22.95 $\pm$ 0.38 &    497928  &    5400& 1.5m-DK\\
R   &        19.02 $\pm$ 0.05 &     37420  &    1800& 1.5m-DK \\
r   &        20.18 $\pm$ 0.01 &     55987  &    1800& LT \\
r   &        20.96 $\pm$ 0.02 &    142301  &    1800& LT\\
r   &        21.72 $\pm$ 0.04 &    216691  &    2100& LT \\
r   &        22.10 $\pm$ 0.09 &    307411  &     780& INT/WFC \\
r   &        22.59 $\pm$ 0.06 &    395798  &    5700& INT/WFC \\

unfilt&	13.70 $\pm$ 0.03&	38 &	5    &	ROTSE-III\\
unfilt&	13.00 $\pm$ 0.03&	52&	5    & ROTSE-III\\
unfilt&	12.79 $\pm$ 0.03&	67&	5    & ROTSE-III\\
unfilt&	12.78 $\pm$ 0.04&	81&	5    & ROTSE-III\\
unfilt&	12.81 $\pm$ 0.03&	95&	5    & ROTSE-III\\
unfilt&	12.63 $\pm$ 0.12&	124&	5    & ROTSE-III\\
unfilt&	12.84 $\pm$ 0.11&	138&	5    & ROTSE-III\\
unfilt&	13.06 $\pm$ 0.19&	152&	5    & ROTSE-III\\
unfilt&	12.94 $\pm$ 0.03&	173&	20     & ROTSE-III\\

\end{tabular}
\caption{Optical data obtained for GRB~080810; errors are at the 1$\sigma$ level. No correction for Galactic extinction has been made. FTS/N = Faulkes Telescope South/North; NOT = Nordic Optical Telescope; OSN = Observatorio de Sierra Nevada; DK = Danish telescope at La Silla; LT = Liverpool Telescope; INT/WFC = Isaac Newton Telescope/Wide Field Camera. KANATA data were taken from Ikejiri et al. (2008). The UVOT $u$, $b$ and $v$ filters are close to the standard $UBV$ system. (Poole et al. 2008). Errors are at the 1$\sigma$ level.}
\end{center}
\end{table}
\addtocounter{table}{-1}
\begin{table}
\begin{center}
\caption{-- {\bf continued}} 
\begin{tabular}{lcccc}
\hline
Filter & Mag. & Time & Exp. time & Telescope\\
 & & (s since burst) & (s)\\
\hline
unfilt&	12.87 $\pm$ 0.19&	261&	20     & ROTSE-III\\
unfilt&	13.65 $\pm$ 0.10 &    386&	20     & ROTSE-III\\
unfilt&	13.76 $\pm$ 0.03&	444&	20     & ROTSE-III\\
unfilt&	13.91 $\pm$ 0.03&	493&	60     & ROTSE-III\\
unfilt&	14.06 $\pm$ 0.04&	562&	60     & ROTSE-III\\

unfilt&	14.14 $\pm$ 0.16&	631&	60     & ROTSE-III\\
unfilt&	14.36 $\pm$ 0.07&	700&	60     & ROTSE-III\\
unfilt&	14.37 $\pm$ 0.07&	770&	60     & ROTSE-III\\
unfilt&	14.49 $\pm$ 0.06&	839&	60     & ROTSE-III\\

unfilt&	14.66 $\pm$ 0.13&	908&	60     & ROTSE-III\\
unfilt&	14.77 $\pm$ 0.14&	977&	60     & ROTSE-III\\
unfilt&	14.70 $\pm$  0.07&	1046&60     & ROTSE-III\\
unfilt&	14.86 $\pm$ 0.04&	1115&60     & ROTSE-III\\

unfilt&	15.16 $\pm$ 0.03&	1504&682 & ROTSE-III\\
unfilt&	15.75 $\pm$ 0.05&	2196& 684&ROTSE-III\\
unfilt&	15.79 $\pm$ 0.1 &   2823& 552 & ROTSE-III\\
unfilt&	16.45 $\pm$ 0.05&	4438&682 & ROTSE-III\\
unfilt&	16.83 $\pm$ 0.05&	5129&682 & ROTSE-III\\
unfilt&	16.94 $\pm$ 0.04&	5821& 682& ROTSE-III\\
unfilt&	17.10 $\pm$  0.06&	6512&682 & ROTSE-III\\
unfilt&	17.29 $\pm$ 0.04&	7203& 682& ROTSE-III\\
unfilt&	17.42 $\pm$ 0.06&	7898&690 & ROTSE-III\\
unfilt&	17.39 $\pm$ 0.15&	9298&	690 & ROTSE-III\\

V & 20.27 $\pm$ 0.07& 49190 & 300 & 1.5m-OSN\\
V & 20.33 $\pm$ 0.08& 51228 & 300 & 1.5m-OSN\\
V & 20.35 $\pm$ 0.09& 53239 & 300 & 1.5m-OSN\\
V & 20.44 $\pm$ 0.11& 54203 & 300 & 1.5m-OSN\\
V   &        13.7 $\pm$ 0.1$^{a}$ &      390   &      33& KANATA \\

v &    13.73 $\pm$ 0.08 &     196 & 10 & UVOT\\
v &    13.81 $\pm$ 0.08 &     206  & 10 & UVOT\\
v &    13.80 $\pm$ 0.08 &     216  & 10 & UVOT\\
v &    13.74 $\pm$ 0.08 &     226  & 10 & UVOT\\
v &    13.89 $\pm$ 0.08 &     236  & 10 & UVOT\\
v &    13.83 $\pm$ 0.08 &     246  & 10 & UVOT\\
v &    13.95 $\pm$ 0.08 &     256  & 10 & UVOT\\
v &    13.74 $\pm$ 0.08 &     266  & 10 & UVOT\\
v &    13.91 $\pm$ 0.08 &     276  & 10 & UVOT\\
v &    14.04 $\pm$ 0.08 &     286  & 10 & UVOT\\
v &    13.98 $\pm$ 0.08 &     296  & 10 & UVOT\\
v &   13.93 $\pm$ 0.08  &    306   & 10 & UVOT\\
v &   14.14 $\pm$ 0.08  &    316   & 10 & UVOT\\

v &   14.00 $\pm$ 0.08  &    326   & 10 & UVOT\\

v &   14.27 $\pm$ 0.09  &    336   & 10 & UVOT\\
v &   14.14 $\pm$ 0.08  &    346   & 10 & UVOT\\
v &   14.09 $\pm$ 0.08  &    356   & 10 & UVOT\\
v &   14.12 $\pm$ 0.08  &    366   & 10 & UVOT\\
v &   14.21 $\pm$ 0.09  &    376   & 10 & UVOT\\
v &   14.28 $\pm$ 0.09  &    386   & 10 & UVOT\\
v &   14.27 $\pm$ 0.09  &    396   & 10 & UVOT\\
v &   14.16 $\pm$ 0.08  &    406   & 10 & UVOT\\
v &   14.49 $\pm$ 0.09  &    416   & 10 & UVOT\\
v &   14.39 $\pm$ 0.09  &    426   & 10 & UVOT\\
v &   14.32 $\pm$ 0.09  &    436   & 10 & UVOT\\
v &   14.46 $\pm$ 0.09  &    446   & 10 & UVOT\\
v &   14.50 $\pm$ 0.10  &    456   & 10 & UVOT\\

v&	17.53 $\pm$ 0.10&	5148&	197&	UVOT\\
v&	21.10 $\pm$ 0.49&	63077&	1079&	UVOT\\
v&	20.79 $\pm$ 0.44&	103938&	777&	UVOT\\

v&	22.05 $\pm$ 0.49&	303764&	5964&	UVOT\\

B & 21.14 $\pm$ 0.12 & 49887 & 300 & 1.5m-OSN\\
B & 21.23 $\pm$ 0.13 & 51901 & 300 & 1.5m-OSN\\

\end{tabular}
$^{a}$ Taken from Ikejiri et al. (2008).
\end{center}
\end{table}
\addtocounter{table}{-1}
\begin{table}
\begin{center}
\caption{-- {\bf continued}} 
\begin{tabular}{lcccc}
\hline
Filter & Mag. & Time & Exp. time & Telescope\\
 & & (s since burst) & (s)\\
\hline
b&	17.89 $\pm$ 0.08&	4532.3&	197&	UVOT\\
b&	20.28 $\pm$ 0.16&	27235&	569&	UVOT\\
b&	21.45 $\pm$ 0.32&	56302&	867&	UVOT\\
b&	21.83 $\pm$ 0.47&	79387&	778&	UVOT\\
b&	21.60 $\pm$ 0.38&102520.4&	831&	UVOT\\
b&	22.55 $\pm$ 0.52&	164312&	2162&	UVOT\\

white & 14.87 $\pm$ 0.10 & 90 & 10 & UVOT\\
white & 14.78 $\pm$ 0.11 & 100 & 10 & UVOT\\
white & 14.68 $\pm$ 0.12 & 110 & 10 & UVOT\\
white & 14.71 $\pm$ 0.12 & 120 & 10 & UVOT\\
white & 14.89 $\pm$ 0.10 & 130 & 10 & UVOT\\
white & 14.83 $\pm$ 0.11 & 140 & 10 & UVOT\\
white & 14.90 $\pm$ 0.10 & 150 & 10 & UVOT\\
white & 14.93 $\pm$ 0.10 & 160 & 10 & UVOT\\
white & 14.93 $\pm$ 0.10 & 170 & 10 & UVOT\\
white & 14.93 $\pm$ 0.10 & 180 & 10 & UVOT\\

\hline
\end{tabular}

\label{optdata}
\end{center}
\end{table}

{\it Swift}-UVOT detected a bright optical afterglow in the $v$-, $b$- and
$white$ filters (Holland \& Page 2008); the non-detection in $u$ and the UV
filters ($\sim$~4~ks after the trigger) is consistent with the redshift of 3.355.

ROTSE-III (Akerlof et al. 2003), at the Siding Spring Observatory in Australia, imaged GRB~080810
35~s after the burst (Rykoff 2008), detecting a counterpart which brightened
for about 30~s, before fading first with a shallow decay, then with a steeper
slope. These unfiltered magnitudes have been normalised to the $R$-band for subsequent analysis; the method is described by Rykoff et al. (2009).

A target-of-opportunity program was triggered on the Keck telescopes
to obtain HIRES spectroscopy (Vogt et al. 1994) of the afterglow of GRB~080810, with the observations beginning around 37~minutes after the trigger (Prochaska et al. 2008); these data showed the redshift of the burst to be z~$\sim$~3.35 (see Section~\ref{zdet}).

The NOT observations (de Ugarte Postigo et al. 2008) began 10.59~hr after the burst, providing
confirmation of the redshift. This was followed by observations with the 1.5~m OSN (Observatorio de
Sierra Nevada; de Ugarte Postigo, Aceituno \& Castro-Tirado
 2008b) telescope and the 1.54~m Danish telescope at La Silla, Chile (Th{\" o}ne, de Ugarte Postigo \& Liebig 2008).

Data from the Faulkes Telescopes North \& South, the Liverpool Telescope (LT; Guidorzi et al. 2008a,b), the Isaac
Newton Telescope (INT/WFC) and the NOT (de Ugarte Postigo et al., 2008a) were reduced in standard fashion, including bias subtraction and flatfielding; $i'$-band observations were additionally defringed. Seeing-matched aperture
photometry was performed using the Image Reduction and Analysis Facility (IRAF)\footnote{IRAF is distributed by the National Optical
Astronomy Observatories, which are operated by the Association of Universities for
Research in Astronomy, Inc., under cooperative agreement with the National Science
Foundation.}. Absolute calibration was performed using the Sloan Digital Sky Survey
(SDSS; Adelman-McCarthy et al. 2008), with conversion to Johnson-Cousins\footnote{following
http://www.sdss.org/dr6/algorithms/\\sdssUBVRITransform.html\#Lupton2005}. 

All the optical data included in this paper are listed in Table~\ref{optdata}, including data from KANATA, obtained from the GCN (Ikejiri et al. 2008). These magnitudes have not been corrected for Galactic extinction (although the fluxes used later have been).

\begin{figure}
\begin{center}
\includegraphics[clip,width=8.5cm]{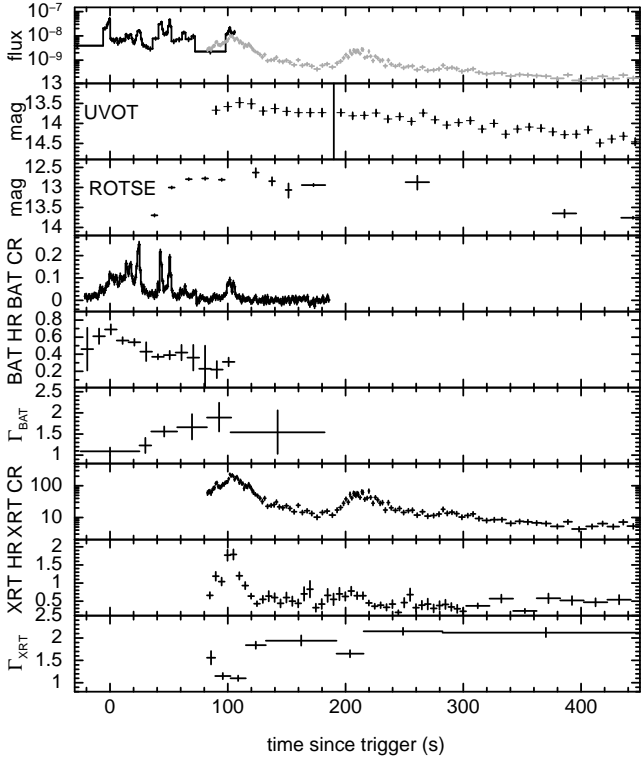}
\caption{From top to bottom: the joint BAT-XRT flux light-curve, in {\bf erg~cm$^{-2}$~s$^{-1}$} over 0.3--10~keV (black histogram -- BAT; grey crosses -- XRT), during the first orbit; the $white$ and $v$-band UVOT light-curve (the data before the vertical line are $white$, normalised to align with the $v$-band; afterwards -- $v$); unfiltered ROTSE-III light-curves; the BAT 15--150~keV light-curve; the BAT hardness ratio over (50--150 keV)/(15--50 keV); time-sliced spectral fits to the BAT data; the XRT light-curve; the XRT hardness ratio over (1.5--10 keV)/(0.3-1.5 keV); time-sliced spectral fits to the XRT data.}
\label{gamma}
\end{center}
\end{figure}

The $white$ and $v$-band UVOT data are plotted in the second
panel of Figure~\ref{gamma}, with the ROTSE-III data in the third. The UVOT white data have been normalised to align with the $v$-band at 180--190~s (shown by the vertical line in the plot). There is an indication that the $white$ magnitude increases around the same time as the first flare seen by the XRT, but this is only significant at the 1.5$\sigma$ level. After this time, the UVOT data appear to follow a smooth decline, with no indication of a brightening at the time of the second X-ray
flare (210~s after the trigger). 

The ROTSE-III data show a clear $\sim$~1~magnitude brightening between T$_0$+40~s and T$_0$+60~s. Although there is apparent structure in the light-curve between
$\sim$120--260~s, the limiting magnitudes were shallow at these times (caused
by cloud cover), meaning these variations are not very significant. 

The lack of obvious flaring in the optical coincident with the X-ray flares implies that this emission
is strongly dominated by the afterglow and has very little (if any) contribution from the prompt
emission.

\section{Discussion}
\label{disc}

In this section, we first present details about how the redshift was determined from the Keck data, then an alternative model for the prompt data and finally discuss the panchromatic observations and what can be learnt from them.

\subsection{Redshift determination}
\label{zdet}

\begin{figure*}
\begin{center}
\includegraphics[clip,width=12cm,angle=90]{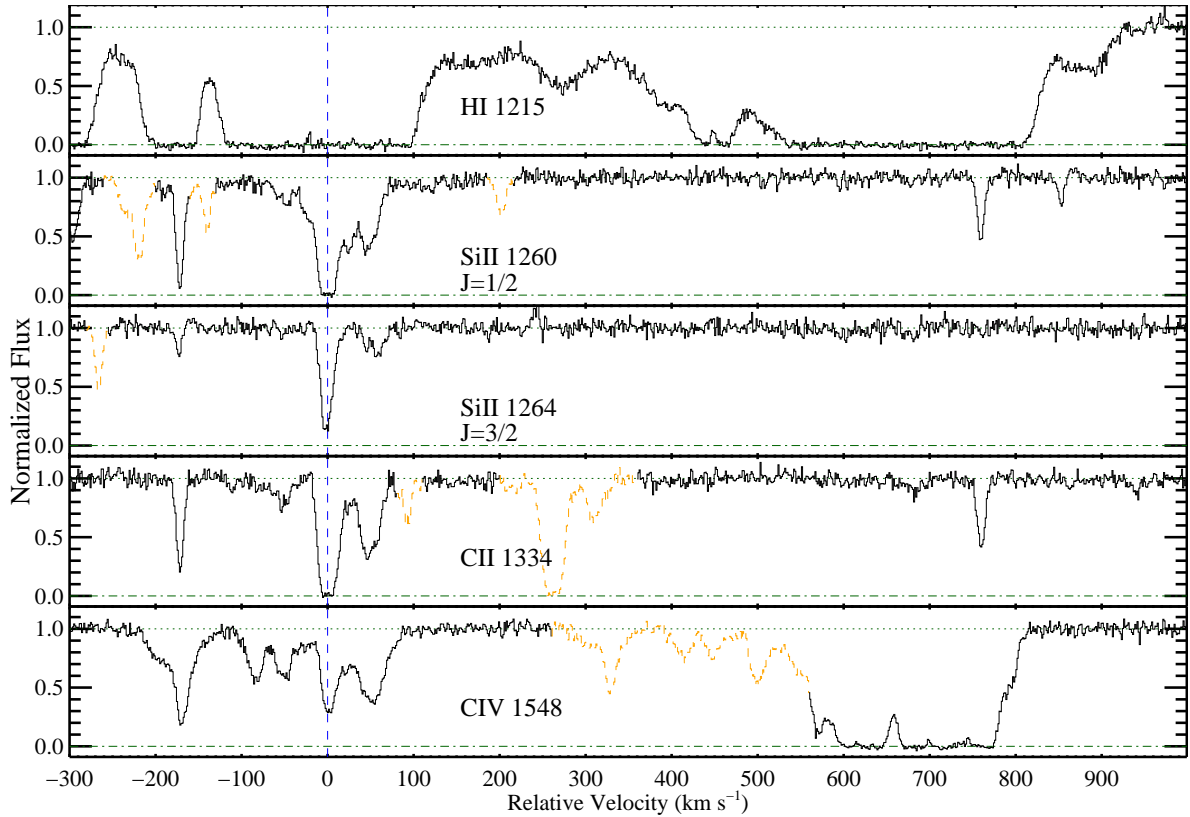}
\caption{Keck/HIRES velocity plots for a subset of the transitions
detected at $z \approx z_{GRB}$.  In the figure v~=~0 km~s$^{-1}$
corresponds to an arbitrary $z=3.35104$ and orange dashed
lines indicate blends with coincident features.  Within the velocity
interval shown, we detect two strong (broad) Ly$\alpha$ lines centred at v~=~0~km~s$^{-1}$
and v $\approx +700$ km~s$^{-1}$  and corresponding metal-line absorption,
including very strong C\,{\sc iv} absorption.  These Ly$\alpha$ lines
are the reddest observed in the afterglow and we conclude that
one (or both) are associated with the GRB host galaxy.
Both complexes also show significant resonance, low-ion absorption from
Si\,{\sc ii} and C\,{\sc ii}  ions, but only the gas at $z \approx 3.35$
shows significant absorption from fine-structure levels of these ions.
}
\label{hires}
\end{center}
\end{figure*}

Figure~\ref{hires} presents a Keck velocity plot for a series of 
transitions corresponding to a $\delta$v $\approx$ 1300 km~s$^{-1}$ interval
around $z \approx 3.355$.  The top panel shows a pair of broad transitions
(centred at v $\approx 0$ and $+$700  km~s$^{-1}$)
which mark the reddest Ly$\alpha$ features detected in the spectrum.  Blueward
of these data are a series of Ly$\alpha$ features which correspond to the
intergalactic medium (IGM) at $z < 3.35$.   The termination of the IGM establishes
a rough redshift (conservatively, a lower limit)
for the GRB host galaxy, i.e. $z_{GRB} \approx 3.355$.

Figure~\ref{hires} reveals that each of these Ly$\alpha$ features also show corresponding 
metal-line transitions of Si\,{\sc ii} , C\,{\sc ii}  (both around 0, 50 and 760~km~s$^{-1}$), and C\,{\sc iv}  (from $-$80 to +80~km~s$^{-1}$ and +580 to almost +800~km~s$^{-1}$).  Also detected is
absorption from O\,{\sc i} , Si\,{\sc iv} , Al\,{\sc ii}  and a number of other ions (not shown in the plot).
However, it is only the gas at $v \approx$ 0 km~s$^{-1}$ (i.e.\ $z \approx 3.351$) 
that shows significant absorption from fine-structure levels of the 
Si\,{\sc ii} , C\,{\sc ii}  and O\,{\sc i}  ions/atoms.   
Absorption from these fine-structure
levels (e.g. Si\,{\sc ii}$^*$ 1264) is a strong signature of gas near the GRB afterglow
because the afterglow itself is the excitation mechanism (Prochaska, Chen \& Bloom 2006).
Therefore, we are inclined to associate the GRB with the redshift of
this material.  Under this hypothesis, the gas at v $\approx$ +700~km~s$^{-1}$
might be considered a neighbouring galaxy with a very large, positive peculiar
velocity.

An alternative hypothesis is that the gas at v $\approx$ +700~km~s$^{-1}$
(i.e., $z \approx 3.36$) marks the ambient interstellar medium of the host
galaxy and that the material at v~$\approx$~0~km~s$^{-1}$ is due to a fast
outflow of material near the GRB afterglow.  This hypothesis is 
challenged by the fast wind speed required and the absence of fine-structure
absorption in the Si\,{\sc ii}  and C\,{\sc ii}  gas at v~$\approx$ +770~km~s$^{-1}$.  
An emission-line measurement of the host galaxy would be required to resolve these two possibilities and so we infer $z_{GRB} = 3.355 \pm 0.005$.

Independent of the correct interpretation of the two systems, 
we can draw a few conclusions
about the gas associated with GRB~080810.  First, the total {\sc hi}
column is low relative to the mean value observed in GRB afterglow
spectra (Jakobsson et al. 2006).  For both Ly$\alpha$
lines we set a conservative upper limit of N$_{\rm HI} < 10^{19.5}$ cm$^{-2}$
based on the absence of strong damping wings.  We also place
a lower limit to the column density of the $z=3.36$ absorber
of N$_{\rm HI} > 10^{17.7}$ cm$^{-2}$ based on Lyman limit absorption at 
$\lambda < 3976$\AA.  Both systems show relatively weak low-ion
absorption, consistent with the low  N$_{\rm HI}$ values. In contrast, each shows very strong high-ion absorption, a characterstic
of gas associated with GRB host galaxies (Prochaska et al. 2007). In agreement with this optical measurement, no absorption above the Galactic value was required when fitting the X-ray spectrum (Section~\ref{sec:x}). 

We note in passing that GRBs~020813 (Barth et al. 2003) and 021004 (Fynbo et al. 2005) both showed similarly unusual absorption systems.

\subsection{Thermal emission}
\label{therm}
Ryde (2005) and Ryde \& Pe'er (2009) discuss the possibility that thermal emission may be ubiquitous in GRBs, fitting samples of spectra with a power-law-plus-blackbody model, and Lazzati et al. (2009) present simulations of photospheric emission. Such thermal components may explain the hard early-time spectra which are inconsistent with synchrotron emission (Preece et al. 1998; Ghisellini, Celotti \& Lazzati 2000; Savchenko \& Neronov 2009). A similar quasithermal model was applied to the GRB~080810 gamma-ray spectra and the results are included in Table~\ref{jointfit}. 
In most of the spectra presented by Ryde \& Pe'er (2009) a distinct break is seen in the temporal evolution of the blackbody temperature; this is not apparent in our data, though this may be related to the fact that our first spectrum spans T$_0$~+~0--10~s, and most of the breaks seen occur within ten seconds of the burst. Following Pe'er et al. (2007), we have estimated the photospheric radius (the radius beyond which the outflow becomes optically thin) to be $\sim$~2~$\times$~10$^{11}$~Y$_0$~cm, where Y$_0$ is the ratio between the fireball energy and the energy emitted in gamma-rays (1$\lo$~Y$_0$~$<$~3--5; see Pe'er et al. 2007). We also determine that the coasting Lorentz factor is $\sim$~570Y$_0^{1/4}$. These calculations assume that the blackbody temperature is the maximum measured (62 keV), since we do not see a break in the behaviour of the temperature. The ratio of the blackbody flux to the total flux over the gamma-ray band (10--10$^{4}$~keV) is 0.2 for this spectrum. With the exception of the spectrum after 100~s, where the blackbody is much weaker and cooler, this ratio remains between 0.1--0.2 for the spectra considered here. We do point out, however, that the fits systematically have a worse $\chi^2$ than the cut-off power-law models.

\subsection{Multi-wavelength data}
\label{multi}

\subsubsection{Broadband modelling}

The top panel of Figure~\ref{gamma} shows the joint BAT-XRT flux light-curve. The data were
converted into 0.3--10~keV fluxes using power-law fits to time-sliced
spectra; where there were simultaneous BAT and XRT detections, the spectra
were fitted jointly. The figure also shows how the BAT and XRT hardness ratios and spectra evolve over time. A general softening trend can be seen across
both bands, though the data harden during the flares, as is typical (e.g., Golenetskii et al. 1983; Ford et al. 1995; Borgonovo \& Ryde 2001;
Goad et al. 2007; Page et al. 2007). This softening with time likely explains why
there was no significant gamma-ray emission detected at the time of the second large
flare seen by the XRT, just after T$_0$+200~s.

The flare around 100~s was detected over both gamma-ray (BAT and GBM) and X-ray bands, though; as mentioned in Section~\ref{sec:opt}, there is only a slight hint of an increase in the UVOT $white$ magnitude at this same time. Extracting and fitting simultaneous BAT and XRT spectra demonstrates that there is curvature between the $\gamma$-ray and X-ray bands; this is also shown by the differences in the independent power-law fits shown in Figure~\ref{gamma}. Broken power-law models are preferred for spectra from both the rising and falling portions of the flare. The break energy is not well-constrained in either case, but is located towards the top end of the XRT bandpass (between $\sim$3--10~keV).

Instead of using a series of broken power-laws to fit GRB light-curves, an alternative method was proposed by
O'Brien et al. (2006) and Willingale et al. (2007): the
light-curve is parametrized by one or two components each comprising an early
exponential rise which then rolls over into a power-law decay. The first of
these components accounts for the gamma-ray and early X-ray emission (the
`prompt' phase), while the second (which is not always seen -- see figure 2 of Willingale et al. 2007) forms what we
typically see as the afterglow at later times. These components allow an
intrinsically curved
light-curve to be fitted without relying on a series of abrupt breaks in a
power-law model. Excluding the times where the flares are seen, this
exponential-to-power-law model can easily account for the BAT-XRT light-curve
of GRB~080810. The second, `afterglow' component begins to dominate in the interval 500--3500~s, between the end of the first and beginning of the second
orbit of data. The optical data can also be modelled with the exponential-to-power-law combination. Using this model, the afterglow E$_{\rm iso}$ was calculated to be 1.6~$\times$~10$^{52}$~erg, approximately a tenth of the prompt E$_{\rm iso}$ (Section~\ref{sec:gamma}). As for the prompt measurement, this is also high, compared with typical values of 10$^{51}$ for afterglows (Willingale et al. 2007).

\begin{figure*}
\begin{center}
\includegraphics[clip,width=10cm,angle=-90]{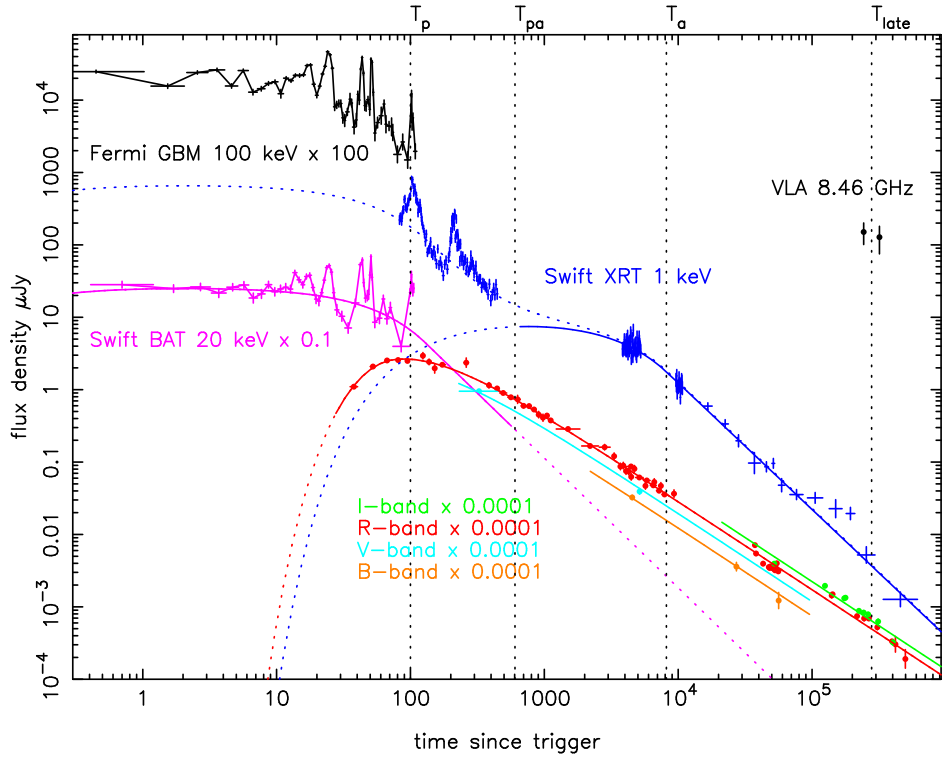}
\caption{Multi-wavelength flux density light-curves, showing the gamma-ray, X-ray, optical and radio data collated for GRB 080810. Four fiducial times are marked by dotted vertical lines and curved lines indicate model fits; both sets of lines are explained in the text. With the exception of the X-ray and radio data, the light-curves have been vertically offset to make the plot clearer; these offsets are given in the figure.}
\label{fdlc}
\end{center}
\end{figure*}

\begin{figure*}
\begin{center}
\includegraphics[clip,width=10cm,angle=-90]{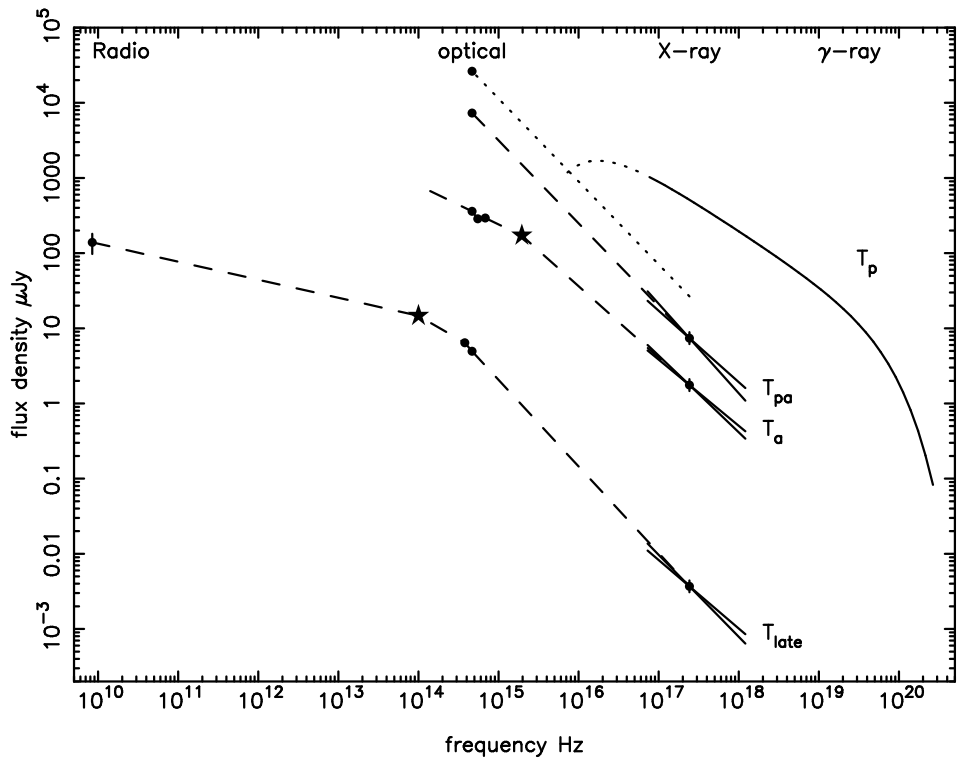}
\caption{Spectral energy distributions of GRB~080810 at the times of the vertical lines in Figure~\ref{fdlc}. The stars indicate the approximate (cooling) break frequencies for the third and fourth SEDs. See text for details.}
\label{sed}
\end{center}
\end{figure*}

Figure~\ref{fdlc} brings together all the photometric data considered in this paper, from the gamma-ray regime to radio wavelengths. The gamma-ray and X-ray light-curves have been converted to flux densities by assuming simple power-law fits (changing over time where there is evidence for spectral evolution). Note that all light-curves except the X-ray and radio points have been vertically offset from their actual positions, to avoid over-lap and make the plot clearer. 

The early high-energy photon index evolves from harder to softer, as indicated by the fact that the Fermi (8--1000~keV data from the summed NaI detectors, normalised to 100~keV) and BAT data extend till $\sim$~100~s, while the softer, 1~keV XRT data show apparent prompt emission out to at least 400~s; at this time {\it Swift} entered the South Atlantic Anomaly and so stopped collecting data for the remainder of the orbit.

The exponential-to-power-law model fits (O'Brien et al. 2006; Willingale et al. 2007) to the BAT, XRT and $R$-band data are shown as solid lines in the figure, with the dotted sections showing the continuation of these models beyond the extent of the data. The model jointly fitted to the BAT-XRT data (with the BAT data extrapolated to 1~keV) is also shown as the higher dotted blue line; the 20~keV magenta line is the same model (though with a different normalisation) as the joint BAT-XRT fit. Each optical band is initially consistent with the same slope ($\alpha_{\rm opt}$~=~1.22~$\pm$~0.09 when the power-law decay takes over), with slightly different normalisations between the filters. Note that the second of the radio points is only a marginal detection, at the 2.4$\sigma$ level. 

The second component fitted to the X-ray and optical profiles also contains an exponential rise. Such an increase is usually assumed because the second component is expected to turn on some time after the prompt trigger; however, in this burst, the early $R$-band points actually show an initial increase, peaking at $\sim$~100~s after the trigger, and provide us with a method of fitting the turn-on of the afterglow. With only three or four points between the initial ROTSE detection and the peak of the optical emission, however, quantifing the speed of the rise would be over-interpreting the data.

We do not have a detailed theoretical model for the onset of the afterglow with which to compare the data. It is possible that part of the rise seen could be caused by spectral evolution (that is, the passage of the peak of the spectrum through the bandpass), although it is reasonable to assume that we are mainly seeing an increase in peak flux as the external (afterglow) shock begins and strengthens, $\sim$~30~s after the burst.

The vertical dotted lines mark four times at which spectral energy distributions (SEDs) have been created (Figure~\ref{sed}): T$_{\rm p}$, at 50~s after the trigger, is during the prompt phase, where Fermi, BAT and ROTSE-III data are available; T$_{\rm pa}$ is the time at which the prompt and afterglow model fits are equal in X-ray flux ($\sim$~600~s after the trigger); T$_{\rm a}$ indicates the time at which the power-law decay starts in the X-ray band ($\sim$~8~ks); T$_{\rm late}$ is the point around which we have X-ray, optical and radio measurements (shortly before $\sim$~300~ks). The optical flux values in Figure~\ref{sed} have been corrected for the Lyman forest absorption (Curran et al. 2008b) and the X-ray data corrected for the hydrogen column. The error bars on the optical points are smaller than the markers plotted.

The first SED, at T$_{\rm p}$, is a distinctly different shape from the later, afterglow-dominated plots. The prompt spectrum is shown as a cut-off power-law (see Table~\ref{jointfit}) with the extension below the X-ray band shown by the curved dotted line, indicating that we assume this component peaks somewhere between the optical and X-ray bands since, as can be seen in Figure~\ref{fdlc}, the optical emission at this time is totally dominated by the afterglow component.
The power-law dotted line from the optical band at this time is parallel to the optical-X-ray slope at T$_{\rm pa}$ -- i.e., it represents the afterglow component. 
There is no evidence for a prompt optical flash (caused by reverse shock emission), since this should be brighter than an extrapolation of the (forward shock) X-ray spectrum. The inferred break between the X-ray and optical bands will be due to synchrotron self-absorption, $\nu_{a}$. Shen \& Zhang (2009) discuss how to constrain the location of the emission region from the self-absorption frequency in a prompt optical-to-$\gamma$-ray SED, assuming the optical and $\gamma$-ray emission are both from the same synchrotron continuum. Their equations give an approximate range of 5~$<$~R~(10$^{14}$$\Gamma_{300}^{3/4}$B$_{4}^{1/4}$~cm)~$<$~16 (for $\nu_{\rm a}$~$>$~$\nu_{\rm opt}$) for the radial distance of the prompt emission from the centre of the GRB explosion. Here, $\Gamma$ is the bulk Lorentz factor ($\Gamma$~=~300~$\times$~$\Gamma_{300}$) and B is the magnetic field strength (B~=~B$_4$~$\times$~10$^4$~G). This distance is of the same order as that estimated by Kumar et al. (2007).

Yost et al. (2007) discuss a number of GRBs where optical observations were obtained during the burst, but prompt optical emission was not detected, as we believe is the case for GRB~080810. They compare the optical-to-gamma-ray spectral indices ($\beta_{opt-\gamma}$) with the gamma-ray spectral slopes ($\beta_{\rm \gamma}$) and find that most of their values imply that there is a spectral rollover between the optical and higher frequencies. The numbers for GRB~080810 at time T$_{\rm p}$ ($\beta_{opt-\gamma}$~$\sim$~0.53 and $\beta_{\rm \gamma}$~$\sim$~0.7) are entirely consistent with their measurements. It seems likely that these events are part of the same population as GRBs with prompt optical detections, but are at the faint end of the distribution.

There is no significant evolution of the X-ray spectral index over the entire period T$_{\rm pa}$ to T$_{\rm late}$, with $\beta_{\rm X}$~=~1.00~$\pm$~0.09. The optical spectral index also remains constant until at least T$_{\rm a}$, with $\beta_{\rm opt}$~=~0.51~$\pm$~0.22 (from a comparison of the $R$, $v$ and $b$ measurements at T$_{\rm a}$). Thus, we find $\Delta \beta$ across the break is $\sim$~0.5 as expected for a cooling break, although the uncertainties on the $\beta$ values are large.

The SED at T$_{\rm pa}$ has been created using only the apparent afterglow flux for the X-ray data -- the lower of the blue lines in Figure~\ref{fdlc}.
At this time, the spectral break must be at or below the optical
band, because the optical flux clearly lies on the direct
extrapolation of the X-ray spectrum.  Between T$_{\rm pa}$ and
T$_{\rm a}$, the X-ray temporal decay is slow, while the optical
has already passed into the power-law phase; that is, the X-ray
decay starts later and is then more rapid than in the
optical [$\alpha_{\rm X}$~=~1.81~$\pm$~0.20 compared to
$\alpha_{\rm opt}$~=~1.22~$\pm$~0.09; we note that, using this method of parametrizing the light-curve, $\alpha_{\rm X}$ is slightly steeper than -- though consistent with -- the value obtained from the series of broken power-laws (see Table~\ref{power})]. Even including the contribution from the residual prompt component (i.e. using the flux from the joint BAT-XRT model -- the upper dotted blue line in Figure~\ref{fdlc}), the X-ray data still decay more gradually than the optical over this time frame.

As an aside, we note that
Granot, Ramirez-Ruiz \& Perna (2005) find that a structured jet
could lead to a steeper decay in the optical than in the X-ray --
the opposite of the measurements for GRB~080810. Using the
standard fireball closure relationships (e.g., Zhang \& M{\'
e}sz{\' a}ros 2004), the optical data at this time are consistent
with a wind medium undergoing slow cooling, with $\nu_{\rm
m}$~$<$~$\nu$~$<$~$\nu_{\rm c}$. In this case,
$\alpha$~=~(3$\beta$+1)/2 =~[(3~$\times$~0.51)~+~1]/2~=~1.27,
compared to the measured value of $\alpha_{\rm opt}$~$\sim$~1.22.

The most straightforward way to explain this difference in the X-ray
and optical behaviour is to have the cooling break moving towards
higher frequencies (i.e. moving from the optical towards the
X-ray). If the blastwave is indeed moving through a wind medium, then
the cooling break is expected to increase with time as t$^{1/2}$. This
would, however, necessitate a transition from a wind to a homogeneous
medium for the break then to move down to lower frequencies (see
later), which would lead to a change in the optical decay which is not
clearly seen.  An alternative explanation for the movement of the
cooling break to higher energies is energy injection. Using the
equations for the luminosity index, $q$, given by Zhang et al. (2006),
we find that this part of the decay is indeed consistent with energy
injection (that is, $q$~$<$~0).
The approximate frequency of this cooling break is marked by a star in
Figure~\ref{sed}; it cannot be determined accurately because of the
lack of constraint on the optical index, but lies within the range
10$^{15}$ -- 2~$\times$~10$^{16}$~Hz.

$BVRI$ data from OSN at 50~ks after the trigger (the closest
measurements were taken and then extrapolated to this exact time
by using the slope of the light-curve) provide a spectral slope
of $\beta_{\rm opt}$~=~0.98~$\pm$~0.38, consistent with the X-ray
data; this confirms that the cooling break has passed through the
optical frequencies by T$_{\rm late}$ and the data are consistent
with the relevant closure relation [$\alpha$~=~(3$\beta$-1)/2 for
$\nu$~$>$~$\nu_{\rm c}$] within the uncertainties. As the cooling
break moves through the optical band, we would expect to see a
change in slope of the light-curve; however, we do not have
sufficient coverage or statistical precision at this late time to
confirm or deny such a break. Furthermore, such a change in slope
is likely to be spread over a decade in time, since the spectral
break is a smooth, rather than abrupt, transition.  With no radio
index and the lack of measurements between the optical and radio
regimes, it is not possible to say where the peak of the spectrum
lies. The star again marks an approximate position for the break,
although we note that there will be multiple breaks in this part of the
spectrum, corresponding to this cooling break, the peak frequency
and the synchrotron self-absorption frequency. he radio data are consistent with the extrapolation of a synchrotron spectrum from the optical and X-ray bands, though.

The optical
spectral measurements are quite poorly constrained, both
at early and late times and, within the errors, $\beta_{\rm opt}$
could be consistent with a constant value from T$_{\rm late}$ to
T$_{\rm a}$.  We also note that the X-ray data during the
power-law decay ($\beta_{\rm X}$~=~1.00~$\pm$~0.09 and
$\alpha_{\rm X}$~=~1.81~$\pm$~0.20) are not consistent with being
above $\nu_{\rm c}$ (specifically, the temporal slope is too
steep and is, in fact, steeper than is typical for a
pre-jet-break slope; see, e.g., Evans et al. 2009), indicating
the decay may not be goverened by the standard fireball model (e.g., Rees \& M{\' e}sz{\' a}ros 1992; M{\' e}sz{\' a}ros \& Rees 1994; M{\' e}sz{\' a}ros \& Rees 1997; Sari, Piran \& Narayan 1998)
with a single synchrotron emission component. This lack of agreement with the standard fireball model appears to be the case for a growing number of afterglows, with many bursts for which we have good, multi-wavelength data presenting challenges which still need to be accounted for (e.g., Willingale et al. 2007; GRB~061007 -- Schady et al. 2007, Mundell et al. 2007; GRB~061121 -- Page et al. 2007).

There is no evidence for a jet break (Rhoads 1999) in the data out to at least six days (which is within the range of {\it Swift} jet break times discussed by Racusin et al. 2009). This places a limit on the jet opening angle of $\theta_{\rm j}$~$>$~4$^o$ (Sari, Piran \& Halpern 1999; Frail et al. 2001) and, hence, a lower limit on the total gamma-ray energy, E$_{\rm \gamma}$~$>$~8~$\times$~10$^{50}$~erg. The beaming fraction, E$_{\rm \gamma}$/E$_{\rm iso}$, is therefore $>$0.0027, which is of the same order as the those found for the samples in Frail et al. (2001) and Bloom, Frail \& Kulkarni (2003).

The so-called canonical XRT light-curve (Nousek et al. 2006; Zhang et
al. 2006) typically starts with a steep decay, with $\alpha$~$\sim$~3, which is
generally attributed to off-axis emission, often referred to as the `curvature effect'
(Fenimore, Madras \& Nayakshin 1996; Kumar \& Panaitescu 2000; Dermer
2004). However, the early X-ray light-curve of GRB~080810 shows a
significantly slower decline, with $\alpha$~$\sim$~1. In fact, the beginning of the GRB~080810 light-curve is very similar to that of GRB~060607A (Ziaeepour et al. 2008; Molinari et al. 2007), both in X-rays and the optical: X-ray flares, superimposed on a relatively slow decline, while the optical emission rises smoothly, independently of the variability seen by the XRT and BAT. Both bursts also show low N$_{\rm HI}$. This slower-than-expected decay could be (at least partly) caused by repeated flaring on top of the underlying continuum. Alternatively, there could be input from the X-ray afterglow; the optical afterglow has clearly risen by this time (with little evidence of prompt emission; Figure~\ref{fdlc}), but the X-ray light-curve cannot be fully deconvolved into the prompt and afterglow components.



No plateau phase is seen in the optical data -- the emission rises and then fades steadily away -- which could be explained by the lack of a strong optical component during the prompt phase: from the exponential-to-power-law model, the plateau phase often seen in the X-ray may be a consequence of the bright prompt emission dominating at early times as the afterglow emission rises (see, e.g., Page et al. 2009 for an example of this). 
The final optical decay starts around 300~s, while the X-ray plateau extends to $\sim$~8~ks before the onset of the power-law decay. These findings (the optical data showing no plateau and beginning the power-law decay phase earlier) were also noted for GRB~061121 (Page et al. 2007), another burst for which good, multi-band data were obtained.

\subsubsection{Onset of the afterglow}

Measurement of the peak time of the afterglow component in the optical band gives us the opportunity to calculate the initial Lorentz factor, $\Gamma_0$
(Molinari et al 2007).
Estimating the peak to occur at $\sim$~100~s (from the ROTSE-III data), and taking E$_{\rm iso}$ to be 3~$\times$~10$^{53}$~erg, $\Gamma_0$ is found to be $\sim$~640($\eta_{0.2}$n$_0$)$^{-1/8}$ (for a constant density medium), at the high end of the range of estimates from other bursts (Molinari et al. 2007; Page et al. 2009; Rykoff et al. 2009). Here, $\eta$ is the radiative efficiency and $n$ is the particle density of the surrounding medium in cm$^{-3}$. However, there is some evidence that the data are better suited to a wind environment, rather than constant density, model. In this case, $\Gamma_0$~$\sim$~235($\eta_{0.2}$)$^{-1/4}$, taking A~=~3~$\times$~10$^{35}$~cm$^{-1}$, where the {\bf number density, $n(r)$~=~Ar$^{-2}$} (Chevalier \& Li 2000; Molinari et al. 2007).

Zhang, Kobayashi \& M{\' e}sz{\' a}ros (2003) and Jin \& Fan (2007) discuss the possible appearance of the reverse shock in optical light-curves. These papers consider three types of reverse shock emission: Type I light-curves show both forward- and reverse-shock emission peaks; Type II show a single peak, corresponding to the reverse shock, with a later flattening caused by the forward shock as it starts to dominate the emission; Type III light-curves show no sign of the reverse shock, with just a single rise and fall of the optical emission. The light-curve of GRB~080810 appears to be a member of this third class, with a single power-law decay seen from around 100~s until several hundred kiloseconds after the burst trigger. The synchrotron frequency, $\nu_{\rm m}$, at the time the reverse shock crosses the outflow is therefore very small, well below the optical band.

Panaitescu \& Vestrand (2008) found an anti-correlation between the peak optical flux and the time of the peak for fast-rising optical afterglows. K-correcting our values to the fiducial redshift of $z$~=~2 used by Panaitescu \& Vestrand gives a peak time of $\sim$~85~s and a corresponding peak flux of $\sim$~94~mJy, meaning that the optical afterglow of GRB~080810 is consistent with their findings. This correlation is explained by a structured outflow being seen off-axis.

Although the rise seen in the optical light-curve is naturally explained as the onset of the afterglow, there are other possible mechanisms -- for example, off-axis or structured outflows or two-component jets (e.g., Granot et al. 2002). These other explanations are discussed for a sample of {\it Swift}-UVOT bursts by Oates et al. (2009). They find that the onset of the forward shock or an off-axis viewing angle could best explain the rises seen in the light-curves.

\subsubsection{Flares}

Prior to {\it Swift}, X-ray observations tended to start hours or days after
the burst trigger, with only a few light-curves showing flaring activity (Piro et al.
1998, 2005). However, flares are now regularly detected by the XRT [see Chincarini et al. (2007) and
Falcone et al. (2007) for survey papers], with about 50\% of {\it Swift}
bursts showing them, typically in the first thousand seconds or so  (Burrows
et al. 2007; Chincarini et al. 2007). Some light-curves, however, do show
rebrightenings at later times (e.g., GRB~050502B -- Falcone et al. 2006; GRB~050724 -- Campana et al. 2006;
GRB~070710A -- Covino et al. 2008; GRB~070311 -- Guidorzi et al. 2007; see also Kocevski, Butler \& Bloom 2007). Confirming the interpretation established by Burrows et al. (2005b) and Zhang et al. (2006), Curran et al. (2008a) investigated late-time flares, finding that
their properties are consistent with those of early ones, implying that the
central engine may sometimes be active for up to 100~ks, or that it
can be restarted. In the case of the deviation from the power-law decay at
about 150~ks in the light-curve of GRB~080810 presented here, the $\Delta$t/t and
$\Delta$F/F values are within the range of the values plotted in figure~3 of
Curran et al. for their sample of late-time flares.  
Because of the uncertainties on the values, this particular
flare could be consistent with internal shocks ($\Delta$t/t~$<$~1), refreshed
shocks or patchy shells (see Ioka, Kobayashi \& Zhang 2005). Both the early and
late flares seen here lie within the distribution of $\Delta$t/t found by
Chincarini et al. (2007).

\section{Summary and conclusions}
\label{conc}

Plentiful broadband data were collected for the bright GRB~080810 which triggered both {\it Swift} and {\it Fermi}. The redshift was found to be 3.355~$\pm$~0.005, and the burst was energetic, with an isotropic energy of 3~$\times$~10$^{53}$~erg in the prompt emission component and 1.6~$\times$~10$^{52}$~erg in the afterglow. There is no evidence for a jet break up to six days after the burst occurred. 

The prompt component (detected from 0.3--10$^{3}$~keV) is seen to evolve from hard to soft, with E$_{\rm peak}$ decreasing from $\sim$~600~keV to $\sim$~40~keV over 110~s. Despite being detected at the same time as the gamma-ray emission, the optical data appear to be strongly dominated by the afterglow, with little (if any) prompt component. SEDs created at fiducial times during the observations show the movement of the cooling break with time. By T$_{\rm a}$ (8~ks after the burst), the break frequency has moved from the optical band towards the X-rays, lying in the range 10$^{15}$ to 2~$\times$~10$^{16}$~Hz; at later times (T$_{\rm late}$~$\sim$~300~ks), this cooling break has evolved to $<$~3~$\times$~10$^{14}$~Hz. Although the optical data conform to the standard fireball interpretation, the decay of the X-ray afterglow (after about 10$^4$~s) is too rapid to be consistent with the model.

Well-sampled bursts such as GRB 080810 enable us to investigate more thoroughly the myriad of models which exist for GRBs, with the ultimate goal of a complete and consistent description of GRB emission from early to late times.

\section{Acknowledgments}

The authors gratefully acknowledge support for this work at the University of
Leicester by STFC, in Italy by funding from ASI and at PSU by NASA contract NAS5-00136. ER thanks the NOVA-3 network for support. J.X.P. is partially supported by NASA/{\it Swift} grant NNX07AE94G. ROTSE-III has been supported by NASA grant NNG-04WC41G and the Australian Research
Council and ESR would like to thank the TABASGO Foundation. A.J.v.d.H. was supported by an appointment to the NASA Postdoctoral Program at the MSFC,
administered by Oak Ridge Associated Universities through a contract with NASA.
F.Y. was supported by NASA Swift Guest Investigator grants NNG-06GI90G and
NNX-07AF02G. The DARK cosmology centre is funded by the DNRF. This paper is partly based on observations made with the Nordic Optical Telescope, operated on the island of La Palma jointly by Denmark, Finland, Iceland, Norway and Sweden, in the Spanish Observatorio del Roque de los Muchachos of the Instituto de Astrofisica de Canarias. This work is partly based on observations with the INT, operated on the island of La Palma by the Isaac
Newton Group in the Spanish Observatorio del Roque de los Muchachos of the Instituto de
Astrofisica de Canarias.
We also thank P. Chandra and D. Frail for help with the radio data, G. Marcey and D. Fischer for scheduling the ToO during their Keck observing time,
Peter Jonker for
performing the INT/WFC observations and C. Th{\" o}ne for working on the data from the NOT and Danish telescope.
We extend our thanks to the whole of the {\it Fermi}-GBM team for their work on this new mission. 
Finally, we thank the anonymous referee for their detailed comments, which improved the paper.




\begin{thebibliography}{}

\bibitem{Adelman-McCarthy08}Adelman-McCarthy J.K. et al., 2008, ApJS, 175, 297
\bibitem{ak03} Akerlof C. et al., 2003, PASP, 115, 132
\bibitem{ar96}Arnaud K.A., 1996, Astronomical Data Analysis Software and Systems V, eds. G. Jacoby and J. Barnes, ASP Conf. Series volume 101, 17
\bibitem{band93}Band D. et al., 1993, ApJ, 413, 281
\bibitem{bar03}Barth A.J. et al., 2003, ApJ, 584, L47
\bibitem{bart05} Barthelmy S.D. et al., 2005, SSRv, 120, 143
\bibitem{bl03}Bloom J.S., Frail D.A., Kulkarni S.R., 2003, ApJ, 594, 674
\bibitem{bo01} Borgonovo L., Ryde F., 2001, ApJ, 548, 770
\bibitem{bu08} Burenin R. et al., 2008, GCN Circ. 8088
\bibitem{bu05a} Burrows D.N. et al., 2005a, SSRv, 120, 165
\bibitem{bu05b} Burrows D.N. et al., 2005b, Science, 309, 1833
\bibitem{bu07} Burrows D.N. et al., 2007, Royal Soc. London Phil. Trans. Series
  A, 365, 1213
\bibitem{ca06} Campana S. et al., 2006, A\&A, 454, 113
\bibitem{ch08} Chandra P. \& Frail, D., 2008, GCN Circ. 8103
\bibitem{chev00}Chevalier R.A., Li Z.-Y., 2000, ApJ, 536, 195
\bibitem{ch07} Chincarini G. et al., 2007, ApJ, 671, 1903
\bibitem{co08} Covino S. et al., 2008, MNRAS, 388, 347
\bibitem{cu08a} Curran P.A., Starling R.L.C., O'Brien P.T., Godet O., van der Horst A.J., Wijers R.A.M.J., 2008a, A\&A, 487, 533
\bibitem{cu08b} Curran, P.A., Wijers, R.A.M.J., Heemskerk M.H.M., Starling R.L.C., Wiersema K., van der Horst A.J., 2008b, A\&A, 490, 1047
\bibitem{de04} Dermer C., 2004, ApJ, 614, 284
\bibitem{deup08a} de Ugarte Postigo A., Th{\" o}ne C.C., Hjorth J., Jakobsson P., Banhidi Z., Grundahl F., Arentoft T., 2008a, GCN Circ. 8089
\bibitem{deup08b} de Ugarte Postigo A., Aceituno F., Castro-Tirado A.J., 2008b,
  GCN Circ. 8090
\bibitem{ev09}Evans P.A. et al., 2009, MNRAS, in press (arXiv:0812.3662v2)
\bibitem{fal06} Falcone A.D. et al., 2006, ApJ, 641, 1010
\bibitem{fal07} Falcone A.D. et al., 2007, ApJ, 671, 1912
\bibitem{fen96} Fenimore E.E., Madras C.D., Nayakshin S., 1996, ApJ, 473, 998
\bibitem{fo95} Ford L.A. et al., 1995, ApJ, 439, 307
\bibitem{fr01}Frail D.A. et al., 2001, ApJ, 562, L55
\bibitem{fyn05}Fynbo J.P.U. et al., 2005, ApJ, 633, 317
\bibitem{gi00}Ghisellini G., Celotti A., Lazzati D., 2000, MNRAS, 313, L1
\bibitem{go07} Goad M.R. et al., 2007, A\&A, 468, 103
\bibitem{gol83} Golenetskii S.V., Mazets E.P., Aptekar R.L., Ilinskii V.N.,
  1983, Nature, 306, 451
\bibitem{gran05}Granot J., Ramirez-Ruiz E., Perna R., 2005, ApJ, 630, 1003
\bibitem{gr07} Grupe D., Nousek J.A., Vanden Berk D.E., Roming P.W.A.,
  Burrows D.N., Godet O., Osborne J., Gehrels N., 2007, AJ, 133, 2216
\bibitem{gr02}Granot J., Panaitescu A., Kumar P., Woosley S.E., 2002, ApJ, 570, L61
\bibitem{gu08a} Guidorzi C., Steele I., Tanvir N., 2008a, GCN Circ. 8093
\bibitem{gu08b} Guidorzi C., Bersier D., Tanvir N., 2008b, GCN Circ. 8099
\bibitem{guid07}Guidorzi C. et al., 2007, A\&A, 474, 793
\bibitem{hol08a} Holland S.T., Page K.L., 2008, GCN Circ. 8095
\bibitem{ik08} Ikejiri Y., Uemura M., Ohsugi T., Kawabata K., Arai A.,
  Yamanaka M., Sakimoto K., 2008, GCN Circ. 8081
\bibitem{io05}Ioka K., Kobayashi S., Zhang, B., 2005, ApJ, 631, 429
\bibitem{jak06} Jakobsson P. et al., 2006, A\&A, 460, L13
\bibitem{jin07}Jin Z.P., Fan Y.Z., 2007, MNRAS, 378, 1043
\bibitem{koc07}Kocevski D., Butler N., Bloom J.S., 2007, ApJ, 667, 1024
\bibitem{ko08} Kocka M., Nekola M., Strobl J., Hudec R., Polasek C., Jelinek
  M., Kubanek P., Munz F., 2008, GCN Circ. 8092
\bibitem{kum00} Kumar P., Panaitescu A., 2000, ApJ, 541, L51
\bibitem{kum07} Kumar P et al., 2007, MNRAS, 376, L57
\bibitem{laz09}Lazzati D., Morsony B.J., Begelman M.C., 2009, ApJL, submitted (arXiv:0904.2779v1)
\bibitem{mal05}Mallozzi R.S., Preece R.D., Briggs M.S. 2005, RMFIT: A Lightcurve
and Spectral Analysis Tool, (Huntsville: Univ. Alabama)
\bibitem{me08} Meegan C.A. et al. 2009, ApJ, submitted
\bibitem{me08} Meegan C.A. et al., 2008, GCN Circ. 8100
\bibitem{mez94}M{\' e}sz{\' a}ros, P., Rees M.J., 1997, ApJ, 476, 232
\bibitem{mez94}M{\' e}sz{\' a}ros, P., Rees M.J., 1994, MNRAS, 269, L41
\bibitem{mol07}Molinari E. et al., 2007, A\&A, 469, L13
\bibitem{mun07} Mundell C.G. et al., 2007, ApJ, 660, 489
\bibitem{no06} Nousek J.A. et al., 2006, ApJ, 642, 389
\bibitem{oa09}Oates  S.R. et al., 2009, MNRAS, 395, 390
\bibitem{ob06} O'Brien P.T. et al., 2006, ApJ, 647, 1213
\bibitem{ok08} Okuma Y., Fujimoto H., Nashimoto T., Wada A., Yonetoku D., Murakami T., 2008, GCN Circ. 8091
\bibitem{pa08} Panaitescu A., Vestrand W.T., 2008, MNRAS, 387, 497 
\bibitem{page07} Page K.L. et al., 2007, ApJ, 663, 1125
\bibitem{page08a} Page K.L. et al., 2008a, GCN Circ. 8080
\bibitem{page08b} Page K.L. et al., 2008b, GCN Report 157
\bibitem{page09} Page K.L. et al., 2009, MNRAS, 395, 328 
\bibitem{pe07} Pe'er A., Ryde F., Wijers R.A.M.J., M{\' e}sz{\' a}ros P., Rees M.J., 2007, ApJ, 664, L1
\bibitem{pi98} Piro L. et al., 1998, A\&A, 331, L41
\bibitem{pi05} Piro L. et al., 2005, ApJ, 623, 314
\bibitem{po08}Poole T.S. et al., 2008, MNRAS, 383, 627
\bibitem{pr98}Preece R.D., Briggs M.S., Mallozzi R.S., Pendleton G.N., Paciesas W.S., Band D.L., 1998, ApJ, 506, L23
\bibitem{pr06} Prochaska J.X., Chen H.-W., Bloom J.S., 2006, ApJ, 648, 95
\bibitem{pr07} Prochaska J.X., Chen H.-W., Dessauges-Zavadsky M., Bloom J.S., 2007, ApJ, 666, 267
\bibitem{pr08} Prochaska J.X., Perley D., Howard A., Chen H.-W., Marcy G.,
  Fischer D., Wilburn C., 2008, GCN Circ. 8083
\bibitem{sav09}Savchenko V., Neronov A., 2009, MNRAS, 396, 935
\bibitem{rac09}Racusin J.L. et al., 2009, ApJ, 698, 43 
\bibitem{rees92}Rees M.J., M{\' e}sz{\' a}ros, P., 1992, MNRAS, 258, 41
\bibitem{rho99}Rhoads J.E., 1999, ApJ, 525, 737
\bibitem{rom06a} Romano P. et al., 2006, A\&A, 456, 917
\bibitem{ro05} Roming P.W.A. et al., 2005, SSRv, 120, 95
\bibitem{ryde04}Ryde F., 2004, ApJ, 614, 827
\bibitem{ryde09}Ryde F., Pe'er A., 2009, ApJ, submitted (arXiv:0811.4135v2)
\bibitem{ry08} Rykoff E.S., 2008, GCN Circ. 8084
\bibitem{ry09} Rykoff E.S. et al., 2009, ApJ, in press (arXiv:0904.0261v1)
\bibitem{sak08a} Sakamoto T. et al., 2008a, GCN Circ. 8082
\bibitem{sak08b} Sakamoto T. et al., 2008b, GCN Circ. 8101
\bibitem{sak09}Sakamoto T. et al., 2009, ApJ, 693, 922
\bibitem{sari99}Sari R., Piran T., Halpern J.P., 1999, ApJ, 519, L17
\bibitem{sari98}Sari R., Piran T., Narayan R., 1998, ApJ, 4997, L17
\bibitem{sch08} Schady P. et al., 2007, MNRAS, 380, 1041
\bibitem{sh09} Shen R.-F., Zhang B., 2009, MNRAS, submitted
\bibitem{stam08}Stamatikos M., Sakamoto T., Band D.L., 2008, in Nanjing Gamma-Ray Burst Conference, Eds. Y.-F. Huang, Z.-G. Dai \& B. Zhang, AIP Conference Proceedings, 1065, 59
\bibitem{star08} Starling R.L.C. et al., 2008, MNRAS, 384, 504
\bibitem{th08} Th{\" o}ne C.C., de Ugarte Postigo A., Liebig C., 2008, GCN Circ. 8106
\bibitem{uem08} Uemura M., Yamanaka M., Ikejiri Y., Sakimoto K., Ohsugi T.,
   Kawabata K., Arai A., 2008, GCN Circ. 8085
\bibitem{vogt94} Vogt S.S. et al., 1994, SPIE, 2198, 362 
\bibitem{vonk04} von Kienlin A. et al., 2004, SPIE, 5488, 763
\bibitem{wil07} Willingale R. et al., 2007, ApJ, 662, 1093
\bibitem{yo08} Yoshida M., Yanagisawa K., Kuroda D., Shimizu Y., Nagayama S., Toda H., Kawai N., 2008, GCN Circ. 8097 
\bibitem{yo07}Yost S.A. et al., 2007, ApJ, 669, 1107
\bibitem{zh03}Zhang B., Kobayashi S., M{\' e}sz{\' a}ros P., 2003, ApJ, 595, 950
\bibitem{zh06} Zhang B., Fan Y.Z., Dyks J., Kobayashi S., M{\' e}sz{\' a}ros P. Burrows D.N., Nousek J.A., Gehrels N., 2006, ApJ, 642, 354
\bibitem{zia08} Ziaeepour H. et al., 2008, MNRAS, 385, 453


\end{thebibliography}
\end{document}